\documentclass{article}
\usepackage[utf8]{inputenc}
\usepackage{graphicx}  
\usepackage{amsmath}   
\usepackage[compress]{cite}
\usepackage{amssymb}   
\usepackage{bm} 
\usepackage{dcolumn}
\usepackage{color,xcolor}
\usepackage{mathrsfs}
\usepackage{amsfonts}
\usepackage{varioref}
\usepackage{braket}
\usepackage{subcaption}
\usepackage{comment}
\usepackage{float}
\RequirePackage[colorlinks,citecolor=blue,urlcolor=magenta,linkcolor=blue]{hyperref}
\labelformat{subsection}{Section #1} 
\labelformat{subsubsection}{Section #1}
\labelformat{subsubsubsection}{Section #1}
\labelformat{figure}{Fig.~#1} 
\labelformat{subfigure}{Fig.~\thefigure#1} 
\labelformat{table}{Tab.~#1} 
\labelformat{appendix}{Appendix #1}
\usepackage{pgfpages}

\usepackage{tikz}   

\usepackage{tikz-3dplot} 

\usepackage{xcolor}
\usetikzlibrary{arrows,positioning} 
\usetikzlibrary{intersections}
\usetikzlibrary{decorations.markings}
\tikzset{
    >=stealth',
    punkt/.style={
           rectangle,
           rounded corners,
           draw=black, very thick,
           text width=6.5em,
           minimum height=2em,
           text centered},
    pil/.style={
           ->,
           thick,
           shorten <=2pt,
           shorten >=2pt,}
}
\usepackage[numbers]{natbib}
\title{When and why do zero-modes cause a divergence in the entanglement entropy?}
\author{Vijay Nenmeli\footnote{vvn21@cam.ac.uk}\quad and\quad  S. Shankaranarayanan\footnote{shanki@iitb.ac.in}
	\\
	\\
{\small{Department of Physics}},
		{\small{Indian Institute of Technology Bombay, Mumbai, 400076, India}}}
\date{ }  

\begin{document}
\maketitle
\begin{abstract}
We examine the correlations between divergences in ground state entanglement entropy and emergent zero-modes of the underlying Hamiltonian in the context of one-dimensional Bosonic and Fermionic chains. Starting with a pair of coupled Bosonic degrees of freedom, we show that zero modes are necessary, but not sufficient for entanglement entropy divergences. We then list sufficient conditions  that identify divergences. Next, we extend our analysis to Bosonic chains, where we demonstrate that zero modes of the entanglement Hamiltonian provide a signature for divergences independent of the entanglement Hamiltonian. We then generalize our results to one-dimensional Fermionic lattices for a chain of staggered Fermions which is a discretized version of the Dirac field. We find that the methods detailed for Bosonic chains have Fermionic analogs and follow this up with a numerical study of the entanglement in the Fermionic chain. Finally, we discuss our results in light of the factorization algebra theorem.
\end{abstract}

\section{Introduction}

Our understanding of physical phenomena from atomic to cosmological scales has vastly improved over the past century. Quantum entanglement is one of these phenomena with possible applications on microscopic and macroscopic scales. Quantum entanglement is a physical quantity with no definite value until it is measured because it is a superposition of all possible outcomes of a hypothetical future measurement~\cite{2009-Horodecki.etal-RMP,Eisert_2010}. In recent years, entanglement has been used to quantify certain physical quantities in quantum systems with many interacting degrees of freedom and strong correlations~\cite{2008-Amico.etal-RMP,2012-Peschel-BJP}. 

Quantum entanglement depends on two properties~\cite{2001-Zanardi-PRL,2011-Thirring.etal-EPJD}: superposition principle and the tensor product structure of quantum states. Since the same state has different tensor product structures in different spaces, the entanglement entropy can be partition-dependent. 
More specifically, quantum states --- represented by density matrices --- are entangled with respect to a tensor product structure in Hilbert space. These are the tensor products of an operator algebra
or observables. Nevertheless, for a given quantum state, one can factorize the algebra that pertains to a density matrix. In the ideal case of complete control of observables, all pure states are equivalent as entanglement resources~\cite{2004-Zanardi.etal-PRL,2011-Harshman.Ranade-PRA}.

While the factorization algebra theorem is powerful, it has limited applicability in two cases. First, it is proven only for finite-dimensional Hilbert space. In other words, the theorem assumes the Hilbert space 
$\mathcal{H}_1 \otimes \mathcal{H}_2$ of operators on the finite dimensional bipartite Hilbert space $\mathcal{H}_1 \otimes \mathcal{H}_2$, with dimension $D=d_1 \times d_2$~\cite{2001-Zanardi-PRL,2004-Zanardi.etal-PRL,2011-Harshman.Ranade-PRA}. Second, it does not provide a definite answer for mixed states. This is because the maximally mixed state is separable for any factorization; thus, a sufficiently small neighborhood surrounding it is also separable. In order to find a factorization that makes the state as entangled as possible, the question becomes how mixed a quantum state can be. 

While early studies of entanglement concentrated on quantum systems with finite dimensions, continuous-variable systems are gaining importance both experimentally and theoretically~\cite{2008-Das.etal-Review,Eisert_2010,2012-Weedbrook.etal-RMP,2013-Braunstein.etal-JHEP}. 
In quantum optics and quantum information processing, for instance, a special class of continuous-variable states, the Gaussian states, play a key role~\cite{2010-Buono.etal-JOpt}. However, entanglement calculations in infinite-dimensional systems are considerably more difficult than their equivalents in finite dimensions.

Besides the analytical evaluation of the reduced density matrix (RDM), the infinite-dimensional systems have an inherent problem --- the divergence of the entanglement entropy. For instance, in the 2-coupled harmonic oscillator (CHO) system for which RDM can be evaluated analytically, the entanglement entropy diverges~\cite{Chandran_2019,2020-Chandran.Shanki-PRD}. While this divergence of entanglement entropy is typically associated with the strong coupling limit, recently, it has been shown that the entropy divergence is due to the emergence of zero modes~\cite{Chandran_2019,2020-Chandran.Shanki-PRD}. More specifically, exploiting an inherent scaling symmetry that the entanglement entropy (RDM) possesses, while the Hamiltonian of the system does not have. Thus, it was shown that entanglement entropy can not distinguish the infinite possible physical systems. 

Interestingly, it has been shown that the emergence of the zero modes is the cause of divergence of entanglement entropy in field theories~\cite{Mallayya_2014,Yazdi_2017,2020-Chandran.Shanki-PRD,2021-Jain.etal-PRD}. This was achieved again through an inherent scaling symmetry of the entanglement entropy that connects the ultra-violet (UV) and the infra-red (IR). Furthermore, it was shown that a scaling symmetry exists in time-independent quantum systems such as the hydrogen atom to quantum fields in curved space-times~\cite{Mallayya_2014,2020-Chandran.Shanki-PRD}. The principal advantage of mapping entropy divergence to the occurrence of zero modes is isolating the divergence component from the non-divergence component. Importantly, the entanglement entropy is insensitive to UV physics in the rescaled variables. Recently, it was shown that an extended scaling symmetry exists for time-dependent systems~\cite{2022-Chandran-Shanki-Arxiv}. 

Given these results, we ask the following questions: Do the presence of zero modes \emph{necessarily} lead to divergence of entanglement entropy? If not, then does it violate the factorization algebra theorem? This work addresses these two questions by first considering CHO. We systematically show that the emergence of zero modes \emph{do not necessarily} lead to divergences in the entanglement
entropy. Specifically, we obtain the following three conditions for CHO:
\begin{enumerate}
\item The zero mode limit can be reached in two ways --- via a limiting free particle Hamiltonian (of an $x^{2}$ Hamiltonian) \textit{or} via an amplitude modulated harmonic oscillator Hamiltonian. The former is associated with divergent entropy, while the latter is not.
\item Entanglement entropy diverges if the ground state associated with the zero mode limit is \textit{non-normalizable}, but stays finite if this state is normalizable.
\item Entanglement entropy diverges if the tower of near zero states approach one another in the {Hilbert space} as the zero mode limit is reached. However, in situations with finite entropy, this cascade of states remains well separated in the Hilbert space, even in the limiting case. 
\end{enumerate}
We then extend the analysis to $(1 + 1)-$dimensional scalar field theories. In this case, we show that while zero modes of the \textit{full} Hamiltonian are not enough to guarantee a divergent entanglement entropy, zero modes of the {entanglement} Hamiltonian \emph{do provide} a necessary and sufficient condition. Finally, we extend the analysis to $(1 + 1)-$dimensional Dirac field. In the case of the Dirac field, the entropy diverges only in the infinite limit and corresponds to the zero modes of the entanglement Hamiltonian. This corresponds to the eigenvalue of the correlation matrix equal to $1/2$. Thus, the entanglement entropy of the Dirac field diverges only in the infinite limit but not in the finite case.

The rest of the article is organized as follows: In Sec. \eqref{sec:CHO}, we probe the origins of entanglement entropy divergences in two-boson systems and study correlations between these divergences and emergent zero-modes. Then, in Sec.~\eqref{sec:Corre-Bosonic}, we extend our analysis to Bosonic chains. Following this, we set up the dynamics of the one-dimensional discretized Dirac Hamiltonian in Sec.~\eqref{sec:FermionSetup}, where we also outline the associated constructs required for an analogous study of the entropy. Next, Sec.~\eqref{sec:FermionAnalytics} details analytic computations of the ground state entanglement entropy for two different partitions of the underlying Hilbert space -- the first involving a two-dimensional subsystem and the second being the infinite subsystem limit. Finally, in Sec.\eqref{Numerics}, we supplant our analytics with numerical computations and briefly go over the connections between zero-modes and diverging entropies for the Fermionic ground state.

\section{Entanglement Entropy with a pair of Bosonic DOF}
\label{sec:CHO}

We begin our analysis by studying quadratic Hamiltonians characterized by two Bosonic degrees of freedom (DOF). As mentioned in the introduction, it is established that such systems exhibit divergences in entanglement entropy even when all numerical parameters describing the system take finite values. Moreover, in specific contexts, these divergences are associated with the emergence of zero modes \cite{Chandran_2019,2020-Chandran.Shanki-PRD}. However, as we show in this section, it is not true that zero modes \textit{necessarily} lead to divergences in the entanglement entropy corresponding to the ground state. 

\subsection{Product States and Zero Modes}

Let us consider the following second quantized Bosonic Hamiltonian, expressed in terms of canonical annihilation operators $\hat{a}$ and $\hat{b}$:
\begin{equation}
    \label{Eg1}
     \hat{H}_{\rm SHO} =\hat{a}^{\dag}\hat{a} + \hat{b}^{\dag}\hat{b} +\lambda (\hat{a}^{\dag}\hat{b}+\hat{b}^{\dag}\hat{a}) \, .
\end{equation}
Under the following canonical transformation $\hat{p}\equiv(\hat{a}+\hat{b})/\sqrt{2}$ and $\hat{q}\equiv(\hat{a}-\hat{b})/\sqrt{2}$, the above Hamiltonian is diagonalized as given by:
\begin{equation}
    \label{Eg1Part2}
   \hat{H}_{\rm SHO} = (1+\lambda)\hat{p}^{\dag}\hat{p}+(1-\lambda)\hat{q}^{\dag}\hat{q}.
\end{equation}
We thus see that the Hamiltonian is bounded below only for $\lambda\leq1$ and has a unique ground state only for $\lambda<1$. Note the emergence of a zero mode as $\lambda\rightarrow 1^{-}$. On the other hand, the ground state is the vacuum $\vert 0\rangle$ of the Fock space, which is, as usual, defined to be the unique state annihilated by the operators $\hat{a}$ and $\hat{b}$ (and hence by $\hat{p}$ and $\hat{q}$). Viewing the Fock space of the complete system as a tensor product of the smaller Fock spaces generated by the operators $\hat{a}$ and $\hat{b}$, it is clear that the ground state is simply a product state in the $\hat{a}/\hat{b}$ basis. That is, $\vert 0\rangle=\vert 0_{\hat{a}}\rangle\bigotimes\vert 0_{\hat{b}}\rangle$, where the states $\vert 0_{\hat{a}}\rangle$ and $\vert 0_{\hat{b}}\rangle$ serve as vacuua for their corresponding annihilation operators. Hence, insofar as the ground state is well defined, its entanglement entropy is never non-zero, let alone divergent. 

The lack of divergence of entanglement entropy as $\lambda$ approaches $1$ can, in this case, be explained using elementary symmetry arguments. To see this, we recall that the most direct way to compute the entropy of a subsystem of a given quantum state is to expand out the state in a ``convenient" basis and trace over the irrelevant degrees of freedom to obtain RDM. In our case, the natural basis to use is, of course, the set of states of constant $``a"$ number and constant $``b"$ number, i.e., states of the form $\vert r,s\rangle$, in the canonical second quantized basis, with $r,s\in\mathbb{Z}^{+}\cup\{0\}$. A generic state can be expressed as $\sum_{r,s}c_{r,s}\vert r,s\rangle$, where $c_{r,s}\in\mathbb{C}\hspace{2pt}\forall\hspace{2pt}r,s$. Thus, a potentially infinite number of non-zero $c$'s could lead to a divergent entropy upon tracing over the $b$ degree of freedom to obtain an RDM for an arbitrary state. However, we are strictly interested in the ground state. Since the total number operator, defined as $N\equiv \hat{a}^{\dag}\hat{a}+\hat{b}^{\dag}\hat{b}$ commutes with the Hamiltonian, and since the ground state is \textit{non-degenerate}, it must be an $N$-eigenstate so that $N\vert\Omega\rangle=n_{\vert\Omega\rangle}\vert\Omega\rangle$ for some non-negative integer $n_{\vert\Omega\rangle}$. As a result, the non-zero $c_{r,s}$s in the ground state basis expansion must necessarily satisfy  $r+s=n_{\vert\Omega\rangle}$. Since only \textit{finitely} many $c$'s can be zero, the density matrix and the resulting RDM are effectively finite-dimensional, leading to a non-divergent entropy. In short, particle conservation constrains the ground state of the Hamiltonian to lie in the span of a \textit{finite} subset of the number eigenbasis, rendering the reduced density matrix, its eigenvalues, and finally, the entanglement entropy finite.

While not incorrect, the above explanation needs to be revised for two reasons: First, it is not generic, and second, it relies on a particular choice of basis. Bogoliubov transformations of the form
\begin{align}
a\rightarrow \lambda a +\mu a^{\dag};\quad
b\rightarrow \lambda b +\mu b^{\dag}
\end{align}
leave the entropy invariant since they do not mix operators defined on the system with those defined on its complement. However, such transformations may well destroy the number-preserving structure of the Hamiltonian so that symmetry arguments of the kind above cannot be applied.

In the rest of this section, we aim to achieve a more precise, yet general, understanding of the connection between zero modes and divergence of the entanglement entropy for $N=2$ systems. 

\subsection{Zero Modes and Entropic Divergences}

Bearing in mind our eventual goals of generalizing to homogeneous Bosonic chains and possibly continua as well, we will demand symmetry of the Hamiltonian under the exchange $\hat{a}\leftrightarrow \hat{b}$. This ``exchange symmetry" is the two DOF remnants of periodic boundary conditions combined with full translation symmetry or, in the case of the Bosonic chain, a discrete lattice symmetry.

The most generic quadratic Hamiltonian we can construct contains combinations of number operators ($\hat{a}^{\dag}\hat{a},\hat{b}^{\dag}\hat{b}$), pair creation/annihilation bilinears ($\hat{a}^{2},\hat{a}^{\dag}\hat{b}^{\dag} etc.$). The number preserving bilinears ($\hat{a}^{\dag}b$) with coefficients suitably constrained to satisfy Hermiticity exchange symmetry and positivity (which ensures that we have a well-defined ground state). In principle, however, we need only demand the weaker condition of a lower bound on energy. Such systems can be easily solved by switching to the analog of a momentum space basis $\hat{p}(\hat{q})\equiv(\hat{a}+(-)\hat{b})/\sqrt{2}$ and (if needed) finding a relevant Bogoliubov transformation to decouple the Hamiltonian. However, as is well known, it is preferable to solve this system by switching to the \textit{real} basis:
\begin{eqnarray}
\label{FockToSHO}
\hat{x}_{1} = \frac{\hat{a}+\hat{a}^{\dag}}{2},  && 
\hat{p}_{1}=\frac{\hat{a}^{\dag}-\hat{a}}{2i}; \quad 
\hat{x}_{2}=\frac{\hat{b}+\hat{b}^{\dag}}{2},
\quad
    \hat{p}_{2}=\frac{\hat{b}^{\dag}-\hat{b}}{2i}, 
\end{eqnarray}
where the generic quadratic Hamiltonian can be thought of as a chain of oscillators. In particular, any Bosonic, quadratic, second quantized Hamiltonian can be rewritten as~\cite{Eisert_2010}:
\begin{equation}
\label{BosonicHamSHO}
\hat{H}_{\rm SHO} =\frac{\hat{p}^{T}P\hat{p}}{2}+\frac{\hat{x}^{T}X \hat{x}}{2} \, ,
\end{equation}
 where $\hat{x}$ and $\hat{p}$ refer to the vector operators $(\hat{x}_{1},\hat{x}_{2})^{T}$ and $(\hat{p}_{1}\hat{p}_{2})^{T}$ respectively and the matrices $X$ and $P$ are positive. For simplicity, we restrict to cases where all the {elements} of these matrices are also positive reals (or possibly zero). The features we will outline will continue to hold in the general case. With these restrictions, $X$ and $P$ can be expressed as four positive real functions $(j, k, l, m)$:
\begin{eqnarray}
\label{MatForm}
X=
\begin{pmatrix}
j&k\\
k&j\\
\end{pmatrix} \, ;
&\quad &
P=
\begin{pmatrix}
l&m\\
m&l\\
\end{pmatrix}  \, .  
\end{eqnarray}
The mode frequencies are 
\begin{equation}
    \label{ModeFrequencies}
\mu_{\pm}=(j\pm k )(l\pm m) \, .
\end{equation}
and the positivity implies $j>k\geq0$ and $l>m\geq0$. 
Hence, zero modes emerge in the following three scenarios: (i) when $j$ and $k$ approach one another, 
(ii) when $l$ and $m$ approach one another, or 
(iii) when the above two scenarios happen simultaneously. We must compute a relevant diagnostic parameter to relate zero modes to entropy. For $N=2$, the symplectic eigenvalue $\alpha$ of the $2\times2$ reduced covariance matrix $C_{R}$ is one such parameter.  $C_{R}$ can be obtained by appropriately truncating the full covariance matrix $C$, which, for a general $N\times N$ harmonic oscillator chain specified by positive matrices $X$ and $P$ (defined above in Eq.~\ref{BosonicHamSHO}), contains information on the correlations in the ground state~\cite{Eisert_2010}:
\begin{equation}
    \label{CovMat}
    C\equiv\begin{pmatrix}
   \Gamma_{x} &0\\
    0&\Gamma_{p}\\
    \end{pmatrix},
\end{equation}
where $\Gamma_{p}=\Gamma_{x}^{-1}=X^{1/2}(X^{1/2}PX^{1/2})^{-1/2}X^{1/2}$.
For $N= 2$, we have: 
\begin{equation}
    \label{SymEig}
    \alpha^{2}=\text{det}(C_{R})=\frac{jl-km}{2\sqrt{(j^2-k^2)(l^2-m^2)}}+\frac{1}{2} \, .
\end{equation}
The entanglement entropy  is related to the symplectic eigenvalue $\alpha$ by the relation~\cite{Mallayya_2014}
\begin{equation}
    \label{SymToEig}
    S=(\alpha+\frac{1}{2})\ln(\alpha+\frac{1}{2})-(\alpha-\frac{1}{2})\ln(\alpha-\frac{1}{2}) \, .
\end{equation}
Thus, the divergences of the entropy may be traced back to divergences in $\alpha$.  

Immediately, we observe that $alpha$ diverges in the three mentioned scenarios --- $j$ and $k$ approach one another, or $l$ and $m$ approach one another or both. Thus, we conclude that zero modes are a \textit{necessary} condition for the divergence. This leads to the following question: Is the presence of zero modes a sufficient condition for the divergence of entropy? The answer to this question, however, is not trivial. In particular, we see that if elements of \textit{only} one of the pairs $(j,k),(l,m)$ approach each other,  the numerator in \eqref{SymEig} stays \textit{non-zero}, so that the entropy \textit{necessarily} diverges. However, if members from each pair converge simultaneously, the numerator vanishes along with the denominator; hence, we must study the resulting limit more carefully.

\begin{figure}[H]
         \centering
\begin{tikzpicture}
\draw[ultra thick,black] (0,0) .. controls (.35,1) and (.75,1.25) .. (1,1.3) ;
  \draw[ultra thick,->,black] (1,1.3) .. controls (1.25,1.35) and (1.6,1.4) .. (2,1.38) ;
  \draw[ultra thick,black] (2,1.38) .. controls (3,1.35) and (4.5,1.32) .. (6,1.3) ;
 \draw[ultra thick,->,red] (0,0) -- (2.15,0.5);
 \draw[ultra thick,red] (2.15,.5) -- (6,1.3);
 \draw[ultra thick,blue] (0,0) .. controls (.35,-0.45) and (.8,-.65) .. (1.15,-.7) ;
  \draw[ultra thick,->,blue] (1.15,-.7) .. controls (1.75,-.71) and (2,-.62) .. (2.1,-.6) ;
  \draw[ultra thick,blue] (2.1,-.6) .. controls (4,.2) and (5,.75) .. (6,1.3) ;
\draw(-0.55,0) node{{\scriptsize \textbf{\Large{(j,k)}}}};
\draw(6.55,1.5) node{{\scriptsize \textbf{\Large{(l,m)}}}};
\draw(1.65,1.75) node{{\scriptsize \textbf{\Large{\textcolor{black}{I}}}}};
\draw(1.8,.9) node{{\scriptsize \textbf{\Large{\textcolor{red}{II}}}}};
\draw(2,-.3) node{{\scriptsize \textbf{\Large{\textcolor{blue}{III}}}}};
\filldraw [black] (0,0) circle (2pt);
\filldraw [black] (6,1.3) circle (2pt);
\end{tikzpicture} 
\caption{{\bf Routes to Degeneracy and Resulting Divergences:}
{While traversing any of the above three paths leads to a Hamiltonian bearing zero modes, the entropy varies wildly among different paths. This is because the entropy remains finite only along the central red curve while diverging logarithmically along the surrounding paths.}}
     \label{ParameterSpacePaths}
\end{figure}
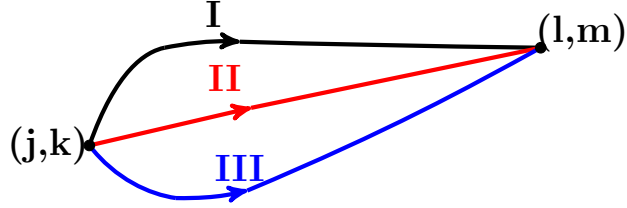

The complexity of the limit stems from its multivariable (and thus directional) character. To see this in action, we consider a system specified by the positive matrices:
\begin{equation}
\label{MatForm2}
X_{0}=
\begin{pmatrix}
j_{0}&k_{0}\\
k_{0}&j_{0}\\
\end{pmatrix} \, ;
\qquad 
P_{0}=
\begin{pmatrix}
l_{0}&m_{0}\\
m_{0}&l_{0}\\
\end{pmatrix}
\end{equation}
and allow \textit{only} the diagonal parameters $j$ and $l$ to vary along a $\tau$ parameterized path so that $(j,l)=(j_{0},l_{0})$ at $\tau=0$ and $(j,l)=(k_{0},m_{0})$ at $\tau=1$. Thus, at $\tau = 0$, the system is non-degenerate and is degenerate at $\tau = 1$. Numerous paths can be traced out in the fictitious 2-dimensional $j-l$ space. \ref{ParameterSpacePaths} contains representative trajectories. It is clear from the onset that the limiting value of the entropy as $\tau\rightarrow 1^{-}$ is highly path dependent. For instance, the path  $(j(\tau),l(\tau))=(j_{0}(1-\tau)+k_{0}\tau,l_{0}(1-\tau)+m_{0}\tau)$ (c.f. curve II in \ref{ParameterSpacePaths}) yields a finite symplectic eigenvalue $\alpha=(ad+bc-2bd)/\sqrt{(a-b)(c-d)}$  in the $\tau\rightarrow 1^{-}$ limit. Thus, the resultant entropy is finite. On the other hand, the asymmetric path $(j(\tau),l(\tau))=(j_{0}(1-\tau)^2+k_{0}(1-(1-\tau)^2),l_{0}(1-\tau)+m_{0}\tau)$ (c.f. Path I in \ref{ParameterSpacePaths} or its $(j,k)\leftrightarrow(k,m)$ equivalent, path III) leads to a diverging eigenvalue $\alpha$ in the $\tau\rightarrow 1^{-}$ limit. Thus, this limit leads to divergence in entropy.

Thus, we have conclusively shown that the entanglement entropy need not compulsorily diverge when both pairs $(j,k)$ and $(l,m)$ approach one another simultaneously. These limits may be more abstractly stated as the limit of $X$ and $P$ approaching non-invertible matrices. The emergence of zero modes is a {necessary} and not a {sufficient} condition for the divergence of the entanglement entropy. This leads us to the following questions: What additional element do we require to constitute a complete, standalone set of requirements for the divergence of the entropy? Furthermore, at a more basal level, {what} is the link between zero modes and highly entangled states? After all, the emergence of zero modes and the resulting tower of states with ``near zero energies" are properties of the {Hamiltonian} and not of the {ground state} per se. The emergence of the zero modes must thus coincide with a drastic change in the structure of the ground state, with the former either serving as an indicator of the latter else directly or indirectly influencing this change. We want to explore such relations in the next subsection. 

\subsection{Geometrical understanding of the cause of divergence}

We may take hints from field theory for this purpose. As is well known, entropy divergences in a  free, $(1 + 1)-$dimensional scalar field theory gain additional IR contributions as we transit from the massive to the massless regime. These extra divergences are accompanied by a loss in normalizability of the ground state \cite{Mallayya_2014,Yazdi_2017,2020-Chandran.Shanki-PRD,2021-Jain.etal-PRD}. While our current system is quantum mechanical, an analogous loss of normalizability emerges in certain degenerate limits, which are accompanied by a divergence in entropy. 

To see this, we examine the Hamiltonian \eqref{BosonicHamSHO}
for possible tell-tale signs. In the decoupled basis $x_{\pm}=(x_{1}\pm x_{2})/\sqrt{2}$, the 2-DOF Hamiltonian \eqref{BosonicHamSHO} splits as
\begin{equation}
    \label{DecoupHam}
    H=H_{+}+H_{-}=\frac{(l+m)p_{+}^{2}}{2}+\frac{(j+k)x_{+}^{2}}{2}+\frac{(l-m)p_{-}^{2}}{2}+\frac{(j-k)x_{-}^{2}}{2}.
\end{equation}
Given our earlier discussions on paths in $j-l$ space, a good way to proceed is to search for discrepancies in the $\tau\rightarrow1^{-}$ limit. One feature immediately sticks out. As we trace along path II in \ref{ParameterSpacePaths}, the decoupled component $H_{-}$ reduces to:
\begin{equation}
\label{HamLimII}
    H_{-}^{\tau\rightarrow 1^{-},\text{II}}=(1-\tau)\Bigl((l_{0}-m_{0})\frac{p_{-}^{2}}{2}+(j_{0}-k_{0})\frac{x_{-}^{2}}{2}\Bigr) \, .
\end{equation}
Along path I in \ref{ParameterSpacePaths}, $H_{-}$ reduces to: 
\begin{equation}
\label{HamLimI}
    H_{-}^{\tau\rightarrow 1^{-},\text{I}}=(1-\tau)\Bigl((l_{0}-m_{0})\frac{p_{-}^{2}}{2}+(1-\tau)(j_{0}-k_{0})\frac{x_{-}^{2}}{2}\Bigr) \, .
\end{equation}
The additional potential prefactor in the latter case implies that $H_{-}$ approaches a (weighted) free particle Hamiltonian before vanishing away completely as $\tau$ approaches 1. This is quite unlike the former, where $H_{-}$ retains the structure of a harmonic oscillator Hamiltonian on its way out. In the $\tau\rightarrow 1^{-}$ limit, this structural dichotomy makes all the difference. 

The emergence of a free particle Hamiltonian in the degenerate limit is not a new observation. In Refs.~\cite{Chandran_2019,2020-Chandran.Shanki-PRD}, this feature was studied in a specific instance of our generic setup, wherein the links between entanglement entropy and the emergence of the free particle Hamiltonian were also elucidated. As per their analysis, the free particle limit can be shown, via a WKB approximation, to yield a translationally invariant RDM, i.e., $\rho(x_{1},x'_{1})=\rho(x_{1}-x'_{1})$. Consequently, correlations in the system do not die out rapidly enough, leading to a divergent entropy. 

The emergence of the free particle Hamiltonian essentially boils down to the differing rates at which the potential and kinetic portions of $H_{-}$ vanish. If, as above, the oscillator potential disappears quicker than the kinetic energy, we proceed to the degeneracy point via a free particle Hamiltonian. On the other hand, if the converse were true, and the potential term faded away more rapidly than its potential counterpart, we reach degeneracy via a rather obscure $H_{-}\rightarrow (1-\tau)x_{-}^{2}$ limit. Nevertheless, the {physics} of the system is unaffected (as evidenced by the canonical transform $x_{-}\rightarrow p_{-},p_{-}\rightarrow -x_{-}$, which regenerates the free particle Hamiltonian of before), and the entanglement entropy diverges once more.

We can remove the divergence in the entropy by varying the potential and kinetic terms at precisely the same rate. In other words, when the rates align perfectly, we approach the degenerate limit via a {harmonic} oscillator Hamiltonian so that the free particle arguments for divergences cease to hold, and the entropy is finite. Thus, zero modes coupled with the emergence of a free particle Hamiltonian (or its position-momentum swapped equivalent) form a sufficient set of conditions for entropic divergences.

We now have one clear-cut method of ascertaining whether or not the entropy will remain finite for a generic quadratic Bosonic system along a generic path in parameter space. That said, an earlier question remains unanswered. Since we concern ourselves only with the ground state, it would be nice to have a diagnostic for divergences that directly involved the ground state rather than, say, knowledge of the limiting Hamiltonian, which is where we currently stand. We reiterate that entanglement entropy, in the 2 DOF case, is a property of the \textit{state}, not the Hamiltonian, so ideally, we would desire a formalism that is fully self-contained to its state vectors.

To understand the nature of the divergences, we must first understand the structure of the ground state in the degenerate limit. The crux of the matter, in the current language, stands thus: The full Hamiltonian, when $l=m$ and $j=k$, is \textit{highly} degenerate, and while all paths in parameter space lead to the $\tau$ parametrized Hamiltonians that converge to a single fixed point, different paths can lead to radically differing ground state limits.

This is a subtle point, and specific examples are the best way to clarify it. Let us consider the variation in the $H_{-}$ ground state as we vary $\tau$ along the example paths in \ref{ParameterSpacePaths}. A precise computation, of course, requires computing the ground state of the full Hamiltonian. However, the regular behavior of $H_{+}$ and the positivity of both $H_{+}$ and $H_{-}$ renders the $H_{+}$ portion of the Hamiltonian irrelevant for present purposes. Curiously, we find that the angular frequency $\omega_{II}=\sqrt{(l_{0}-m_{0})(j_{0}-k_{0})}$ and the ground state of $H_{-}$ remain the same along path II. The $\tau$ factors manifest as an \textit{overall} scale factor for $H_{-}$, but this does not affect whatsoever its eigenstates (and in particular the ground state). The eigenvalues undergo a rescaling, but that is not of concern yet. This provides another way of interpreting the lack of divergence along this path --- knowledge of the ground state alone is insufficient to determine our position on curve II. (Formally, the knowledge of the state vector is not enough to return an exact $\tau$ value corresponding to one's position along the curve). Since the entropy in the 2 DOF case depends solely on the state vector, it follows that it is also insufficient for determining our position on curve II. Also, since the entropy is the same along the curve, the entropy does not diverge along curve II. 
In an informal sense, the $\tau$ prefactor resembles the amplitude one encounters when dealing with the classical oscillator, where a change in the amplitude is akin to spatial rescaling but does not otherwise affect the dynamics. 

On the other hand, path I presents a non-trivial limit, with the \textit{potential} prefactor in Eq.~\eqref{HamLimI} altering the effective \textit{frequency} of the resulting harmonic system. The ground state remains a harmonic oscillator eigenstate, but with an altered, $\tau$ dependent frequency $\omega_{I}(\tau)=\sqrt{(1-\tau)(l_{0}-m_{0})(j_{0}-k_{0})}=\omega_{II} \sqrt{1-\tau}$. Functionally, in the $\tau\rightarrow 1^{-}$, the \textit{normalized} ground state wave function converges pointwise to the zero function and is {not normalizable}. This is the QFT analog that we alluded to earlier. As in the scalar field case, IR divergences of entanglement entropy are signaled by a singular loss of normalizability of the ground state. We have thus identified a sufficient \textit{state} dependent diagnostic of the divergences encountered.

A final diagnostic, one that will become important later on, can be obtained by extending our previous approach to encompass not just the ground state but the tower of excited states obtained by repeatedly acting $\hat{a}_{+}^{\dag}$ on the ground state. 
As we traverse arbitrary paths through parameter space, this tower of states changes continuously.
The behavior of the states of this tower in the (path-dependent) degenerate limit can be used as a signal for diverging entropy. Specifically, from Eq.~\eqref{HamLimII}, it is clear that the tower of states does not alter form as we move along path $II$. Explicitly, the tower of states is simply the set of eigenstates of an SHO with frequency $\omega_{II}$, regardless of the position on the curve. Thus, the tower of states stays mostly the same as we approach the singular limit, which ties in with the absence of divergence along this path. However, along path $I$, the frequency of this oscillator varies with $\tau$ as $\omega_{1}(\tau)$. This variation is non-trivial in the degenerate limit and is a signature of the associated entropy divergence. This non-trivial behavior can easily be seen in the position basis, where \textit{all} the corresponding wave functions flatten out to zero in the degenerate limit. As a result, the tower of near-zero eigenstates \textit{collapses} in this degenerate limit, and this collapse signifies an oncoming divergence. 

In the above discussion, we have listed several features that demarcate the situations that do and do \textit{not} lead to the divergence of entanglement entropy. We enumerate them once more below for clarity.
\begin{enumerate}
\item The zero mode limit can be reached in two ways --- via a limiting free particle Hamiltonian (of an $x^{2}$ Hamiltonian) \textit{or} via an amplitude modulated harmonic oscillator Hamiltonian. The former is associated with divergent entropy, while the latter is not.
\item The entanglement entropy diverges if the ground state associated with the zero mode limit is \textit{non-normalizable}, but stays finite if this state is normalizable.
\item The entanglement entropy diverges if the tower of near zero states `collapses' (at least in the sense of the corresponding wavefunctions flattening out to zero in a naively taken limit.)
\end{enumerate}
For now, we have heuristically established that any one of these features --- free particle Hamiltonians, non-normalizable limits, and a "Cauchy" tower of states --- implies the other. Both of them, in conjunction with the requirement of zero-mode frequencies, provide us with acceptable conditions for a zero-mode divergence. In the following section, we extend the analysis to N-harmonic chains by relating the entanglement Hamiltonian and the covariance matrix~\cite{Peschel_2003}. 

\section{Zero modes and entanglement in Bosonic Chains}
\label{sec:Corre-Bosonic}

The 2-Boson system was an ideal toy model to get started. Small enough to admit explicit expressions for the entropy, it was nevertheless complex enough to allow for divergences in the entanglement entropy. We expect that the conceptual framework understanding divergences is still the same as we move from $2$ to $N$-Bosons. That is, our requirements for divergent entropy --- zero modes and a suitably generalized form of the three points we had listed in the preceding section --- stay the same. The first of these requirements, zero modes, was, in fact, previously proved only in a restricted context~\cite{Mallayya_2014,2020-Chandran.Shanki-PRD}. After all, we assumed that our workhorse 2-Boson system contained only real parameters and posited that the extensions posed no problems. A justification for this statement, along with a proof for the generic $N$ Boson case, comes from the introduction of an intermediary --- the \textit{entanglement Hamiltonian}~\cite{Peschel_2003,2021-Jain.etal-PRD}. 

As demonstrated in Ref.~\cite{Peschel_2003}, the RDM ($\rho$) for a generic subsystem of an integrable 1-D Bosonic chain can be expressed as
\begin{equation}
    \label{EHDefn}
    \rho=\mathcal{K}e^{-\mathcal{H}} \, ,
\end{equation}
where $\mathcal{K}$ is a constant and $\mathcal{H}$ is the \textit{entanglement Hamiltonian} that is quadratic in the Fock space operators of the \textit{subsystem}. In particular, with $a$ and $b$ representing subsystem indices, the entanglement Hamiltonian may be expressed as
\begin{equation}
\label{EHDefn2}
\mathcal{H}= \frac{p^{T}\mathcal{P}p}{2}+\frac{x\mathcal{V}x}{2}
\end{equation}
where $\mathcal{P}$ and $\mathcal{V}$ are positive definite matrices, and the vectors $x$ and $p$ are now confined to the \textit{subsystem} variables. For a given subsystem length $L$, the mode frequencies ($\mu_{a}$) of the entanglement Hamiltonian ($\mathcal{H}$) are related to the (strictly positive) eigenvalues $\nu_{a}$ of the matrix $M\equiv \Gamma_{x}\Gamma_{p}$ by the relation:
\begin{equation}
    \label{CovMatToEntHam}
\frac{\nu_{a}}{2}=\pm \text{coth}\left(\frac{\mu^{*}_{a}}{2}\right) \, .
\end{equation}
Note that $M$ is the product of the position and momentum blocks of the full covariance matrix $\Gamma$.

Using these simple relations, we may derive stringent conditions for the divergence of entanglement entropy as follows - such divergences necessarily require the divergence of a \textit{symplectic} eigenvalues of the full covariance matrix $\Gamma=\Gamma_{x}\oplus\Gamma_{p}$. Now, a standard result from linear algebra tells us that the set of  symplectic eigenvalues of a $2K\times2K$ positive matrix $M$ is simply the set of \textit{positive} eigenvalues of the matrix $i\Omega M$~\cite{Eisert_2010} where $\Omega$ is the standard symplectic form: 
\begin{equation}
    \label{SymFormDefn}
    \Omega=\begin{pmatrix}
    0_{K}&&\mathbb{I}_{K}\\
    -\mathbb{I}_{K}&&0_{K}\\
    \end{pmatrix}
\end{equation}
It is here that the block diagonal nature of the covariance matrix helps us, for we may thus use the previous result to obtain the symplectic eigenvalues of the covariance matrix from the usual eigendecomposition of the matrix
\begin{equation}
    \label{SymEigCovMatStp}
   i\Omega\Gamma= \begin{pmatrix}
      0_{K}&&P\\
    -X&&0_{K}\\
   \end{pmatrix}
\end{equation}
whose characteristic polynomial is easily computed to be $\vert\lambda^{2}\mathbb{I}_{K}-XP\vert=\vert\lambda\mathbb{I}_{K}-XP\vert\vert\lambda\mathbb{I}_{K}+XP\vert=0$ so that the symplectic values of the covariance matrix are {exactly} the numbers $\nu_{a}$. From Eq.~\eqref{CovMatToEntHam}, it is then clear that the symplectic eigenvalues $\nu_{a}$ diverge \textit{only} when the corresponding $\mu_{a}^{*}$s are zero --- when the {entanglement} Hamiltonian develops zero modes. A simple rephrasing of this argument shows that the {converse} is also true - symplectic eigenvalues diverge \textit{only} when zero modes emerge for the entanglement Hamiltonian. 

We can draw additional parallels between zero modes of the Hamiltonian and those of the {entanglement} Hamiltonian. To see this, note that divergent symplectic eigenvalues indicate the divergence of at least one of the {matrix elements} of the covariance matrix. From the definition of the covariance matrix and the above results, it is thus clear that zero-modes of the entanglement Hamiltonian beget divergences of one or more correlation functions involving the position operators $\hat{x}_{i}$ and momentum operators $\hat{p}_{i}$ living in the subsystem in question. Confining ourselves to position correlations, we can rewrite arbitrary two-point functions $\langle\hat{x}_{i} \hat{x}_{j}\rangle$ as linear combinations of two-point functions $\langle\hat{x}'_{i}\hat{x}'_{j}\rangle$ involving the decoupled position coordinates $x'$. We can evaluate this last set of two-point functions by expressing the decoupled coordinate operators $\hat{x}'$ in terms of the ladder operators and their canonical commutation relations. Now, the relation $x'_{i}=(a'_{i}+a_{i}^{'\dag})/\sqrt{2\omega_{i}}$ (where $\omega_{i}$ is the mode frequency and $a_{i}$ is the lowering operator corresponding to the $i^{th}$ decoupled coordinate) implies that transiting from a given decoupled coordinate $x'_{i}$ to its corresponding ladder operators introduces factors of $\omega_{i}^{-1/2}$. Going over the outlined procedure thoroughly, we see those divergences of two-point functions can {only} originate from the frequency factors $\omega_{i}^{-1/2}$ since all other steps of the procedure strictly involve finite coefficients. We can prescribe analogous steps for correlation functions between momentum operators. The key difference is that the frequency factors, in this case, take the form $\omega_{i}^{1/2}$ and are thus finite, even in the zero-mode limit. 

To summarize, we argue that zero-modes of the entanglement Hamiltonian imply divergence of at least one two-point function. Moreover, the frequency factors tell us that the two-point functions can diverge only if (i) they involve positions and not momenta and (ii) at least one of the mode frequencies of the Hamiltonian approaches zero. As a result, we thus see that \textit{zero-modes of the entanglement Hamiltonian necessarily lead to zero-modes of the Bosonic chain Hamiltonian}. This key result allows us to solidify the remarks made in the previous section and tie them to our analysis of Bosonic chains. Specifically, using this result and the equivalence between zero-modes of the entanglement Hamiltonian and entropic divergences, we can verify the claim made earlier: That zero modes of the system Hamiltonian are \textit{necessary}, but not \textit{sufficient} for producing divergences of the entanglement entropy. In particular, only zero modes of the system Hamiltonian that generate zero modes for the corresponding entanglement Hamiltonian lead to divergences of the entanglement entropy.

All our discussions thus far have strictly involved Bosonic degrees of freedom. While Fermionic systems may lead to entanglement entropy divergences, the general low dimensionality of the underlying Hilbert spaces --- stemming from the finite size of a Hilbert space comprising a {single} Fermionic degree of freedom --- means that simple models analogous to those discussed above may {never} show a divergence in the entanglement entropy, zero modes or not. As such, we have to directly begin with infinite degrees of freedom to even \textit{obtain} divergences. Understanding these divergences comes only after that. With this in mind, we will, in the next section, aim to posit some analogous results for the Fermionic case. Our workhorse will be a particularly relevant model for field theory --- the discretized Dirac field in $(1+1)-$dimensions.

\section{Correlation matrix for the Fermion Field}
\label{sec:FermionSetup}
As is commonly done in Bosonic field theory, we will begin our study of entanglement in Fermionic fields by discretizing the Fermionic Hamiltonian. However, the discretization of the Fermionic Hamiltonian is more subtle than its Bosonic counterpart, and we must thus go about this with care~\cite{NIELSEN1981219,Stacey_1982,1975-Kogut.Susskind-PRD,Susskind_1977,Pacholski_2021}. In this section, we briefly discuss the discretization procedure of the Dirac field in $(1 + 1)-$D and then evaluate the correlation matrix.

\subsection{Discretizing the Fermion field}

Crudely put, discretizing a continuum Hamiltonian means defining a lattice model which, in a suitable continuum limit, yields the same physics as the former. Note that this means that the discretized Hamiltonian must be Hermitian. In the Bosonic case, discretization is straightforward. Beginning with the Klein-Gordon Hamiltonian, we first replace the continuum fields and their conjugate momenta with suitably renormalized discrete fields. Second, introduce a lattice parameter $a$, and replace integrals by summations and derivatives by forward differences. Renormalization is required as a naive substitution would transform the continuum canonical commutation relations (CCR) $[\phi(x),\phi(y)]=i\delta(x-y)$ to $[\phi(ma),\phi(na)]=i\delta_{mn}/a$~\cite{montvay_munster_1994,Peskin_1995,smit_2002,Srednicki_2007}. Therefore, the discrete analogs of the continuum CCR are {not} in canonical form. While taken for granted, this naive discretization procedure is perfectly valid since the resulting Hamiltonian is Hermitian for all $a$ and, while non-trivial, can be shown to reproduce the physics of the continuum in the $a\rightarrow 0$ continuum limit. A simple illustration of the continuum limit is that the spectrum of the discrete Hamiltonian (in the relativistic quantum mechanics (RQM)) yields the relativistic dispersion relations in the $a\rightarrow 0$ limit.

Now, the continuum Dirac Hamiltonian in $(1+1)$-D can be obtained by the usual RQM procedure --- we construct the relevant Dirac operator by finding a set of two $2\times 2$ Hermitian matrices $M_{1}, M_{2}$ that satisfy $\{M_{i}, M_{j}\}=2 \delta_{ij}, i,j\in{1,2}$. The obvious choices for the $M$ matrices are the Pauli matrices or combinations thereof. There are infinitely many possibilities; however, like in $(3+1)-$D, they can all be unitarily equivalent~\cite{montvay_munster_1994,Peskin_1995,smit_2002,Srednicki_2007}. We may thus pick an arbitrary choice. The Fermionic Hamiltonian in  $(1+1)$-D is then given by
\begin{equation}
    \label{FermHam1+1D}
    H_{D}=\int{dx \left[-i\psi^{\dag}\alpha\partial_{x}{\psi}+m\psi^{\dag}\beta\psi\right]}
\end{equation}
where $\alpha$ and $\beta$ are the generators of the Clifford algebra mentioned above. In the literature, one chooses $\alpha=\sigma_{2}$ and $\beta=\sigma_{3}$ as this leads to a Hamiltonian with manifestly real coupling coefficients. Although other choices will prove superior later, we will persist with this option.

We note that the above Hamiltonian is \emph{linear} in spatial derivatives (unlike the Bosonic case, which is \emph{quadratic}) and that the first bilinear comprises two fields arranged in an {asymmetric} fashion, with {only one} of the bilinears acted on by a derivative operator. These two features make the discretization of the above Hamiltonian non-trivial. In particular, the usual forward difference method fails as the resulting discrete Hamiltonian is \textit{non-Hermitian}. Moreover, the obvious alternative of approximating the derivative by {symmetric} differences also fails as the spectrum of the resulting Hamiltonian does {not} match the continuum spectrum ($a\rightarrow 0$ limit) --- this is known as the fermion doubling problem \cite{1975-Kogut.Susskind-PRD,Susskind_1977}. Mathematically, the failure of the discrete model to yield the correct spectrum in the continuum limit can be mapped to the fact that the continuum field equations, while first order, yield discrete second-order algebraic equations. In other words, we must specify the initial values of the field at {two} lattice points to obtain a unique solution.

There are several solutions and discretization schemes which bypass the issues mentioned above. In this work, we consider the method of staggered fermions~\cite{1975-Kogut.Susskind-PRD,Susskind_1977}, a scheme prominently applied in lattice studies. The staggered Fermion method in $(1+1)-$D involves defining two sublattices, symmetrically intercalated. The top and bottom components of the continuum Weyl fields are modeled by lattice variables defined on alternating odd and even lattice sites, respectively. One could also regard this setup as one involving two separate Fermionic fields, one defined over odd lattice sites and the other over even sites. However, it must be noted that this point of view fails to manifest the Lorentz transformation properties of the Weyl field. Nevertheless, the resulting massless Hamiltonian can yield the correct spectrum, in the RQM sense, in the continuum limit. 
The discretized {massless} Hamiltonian, from the former viewpoint, is simply
\begin{equation}
    \label{FermHam1+1DLatSussMass0}
    H_{0}=\frac{i}{2a}\sum_{i=1}^{L}\left[\phi^{\dag}(n)\phi(n+1)-\phi^{\dag}(n+1)\phi(n)\right]
\end{equation}
where $\phi$ is a discrete \textit{one component} Fermionic field whose values at even(odd) sites model the first(second) components of the continuum Weyl spinor. Note that $L$ must therefore be an even integer. $\phi$ is a canonically conjugate variable in that
\begin{equation}
    \label{CCRSuss}
    \{\phi(n),\phi^{\dag}(m)\}=\delta_{nm} \, ; \qquad 
    \{\phi(n),\phi(m)\}=0
\end{equation}
The resulting massless Hamiltonian yield the relativistic energy-momentum dispersion in the continuum limit. Mass terms can be added using the $\beta$ matrix from the Clifford algebra. In our convention, with $\beta = \sigma_{3}$, mass terms appear in the Hamiltonian as
\begin{equation}
\label{FermHam1+1DLatSussMassTerm}
   H_{M}= \sum_{i=1}^{L} (-1)^{i}m\phi^{\dag}(n)\phi(n) \, ,
\end{equation}
with the $(-1)^{i}$ coming from the negativity of the $(2,2)$ component of the $\sigma_{3}$ matrix. The full Hamiltonian is thus given by:
\begin{equation}
\label{FermHam1+1DLatSussFull}
 H=H_{0}+H_{M}=\sum_{i=1}^{L} \left[\frac{i}{2a}(\phi^{\dag}(n)\phi(n+1)-\phi^{\dag}(n+1)\phi(n))+(-1)^{i}m\phi^{\dag}(n)\phi(n) \right] \, .
\end{equation}
In Appendix \ref{app:spectrum}, we evaluate the spectrum and eigenstates of the above Hamiltonian on a ring with $L=2N$ points with periodic boundary conditions. In the rest of this section, we evaluate the real space entanglement entropy using the correlation matrix approach using for Bosonic spin chains in \ref{sec:Corre-Bosonic}.

\subsection{The correlation matrix}

As stated before, real space entanglement entropy computations for spin chains usually proceed via evaluation of the matrix of two-point correlators $C$, defined by the second quantized operators $q_{i}$:
\begin{equation}
\label{CMatrixDefn}
C_{ij}=\langle q_{i}^{\dag}q_{j}\rangle \, ,
\end{equation}
where $\langle\mathcal{O}\rangle$ denotes the vacuum expectation value (VEV) of the operator $\mathcal{O}$. The entanglement entropy emerges from the eigenvalues $\lambda_{i}$ of the matrix $C$ as \cite{Latorre_2009}:
\begin{equation}
    \label{CorToEE}
    S=\sum_{i} (1+\lambda_{i})\ln(1+\lambda_{i})+(1-\lambda_{i})\ln(1-\lambda_{i}) \, .
\end{equation}
Such expressions are real only when the correlation matrix possesses eigenvalues in the interval $[0,1]$ with the endpoints accommodated by limits. In fact, it is at least true for our case that if $\lambda$ is an eigenvalue, then so is $1-\lambda$. One should contrast this with the Bosonic analog of correlation matrix eigenvalues --- {symplectic} eigenvalues of the \textit{covariance} matrix. In general, neither of these two quantities --- covariance matrix elements or their eigenvalues are required to obey any such constraints. So, we see that divergences in the entanglement entropy of Fermionic systems can only manifest as a divergence in the {sum}  rather than in the {individual} elements of the {series} \eqref{CorToEE}. This is unlike the Bosonic case, where the argument of the logarithms can themselves diverge. This is a mathematical reflection of the physical statement that Bosonic systems require just \textit{one} zero mode to source a tower of near-zero energy states. In contrast, Fermionic systems require infinitely many zero modes or near zero modes to do so. 

Let us now evaluate the correlation matrix for a subset of $2K$ lattice points. See Appendix \eqref{app:spectrum} for detailed calculations. We have three kinds of correlations ---  correlation between two $d$'s, correlation between the two $b$'s, and the correlation between  $d$'s and $b$'s. We evaluate these individually before combining them. Using the anti-commutation relations between the momentum space ladder operators \eqref{kSpaceOpsDefn},\eqref{RotOpskSpace} and the fact that a pair of these  ladder operators annihilate the vacuum (See Appendix \eqref{app:spectrum}), we obtain the following VEVs in momentum space:
\begin{equation}
    \label{CMatrixKSpaceRot}
\langle\tilde{a}^{\dag}(k)\tilde{a}(k')\rangle=\langle\tilde{c}^{\dag}(k)\tilde{a}(k')\rangle=\langle\tilde{a}^{\dag}(k)\tilde{c}(k')\rangle=0,~~ \langle\tilde{c}^{\dag}(k)\tilde{c}(k')\rangle=\delta_{kk'} \, .
\end{equation}
Transforming to $\tilde{b}$'s and $\tilde{d}$'s using \eqref{RotOpskSpace}, we get 
\begin{equation}
    \label{CMatrixKSpace}
\langle\tilde{b}^{\dag}(k)\tilde{a}(k')\rangle=\vert\beta_{k}\vert^{2}\delta_{kk'};~~
\langle\tilde{d}^{\dag}(k)\tilde{d}(k')\rangle=\vert\alpha_{k}\vert^{2}\delta_{kk'};~~\langle\tilde{b}^{\dag}(k)\tilde{d}(k')\rangle=-\alpha_{k}\beta_{k}^{*}\delta_{kk'}
\end{equation}
where $\alpha_{k}$ and $\beta_{k}$ are defined as
\begin{equation}
\label{AlphaBetaDefn}
\alpha_{k}=\frac{1}{\sqrt{\csc ^2(a k) D^2+1}};~~
\beta_{k}= { (i - \cot (a k))} \, \alpha_{k} \, ; \quad D = (\sqrt{a^2 m^2 - \sin^2 (a k)}- a m)/2 \, .
\end{equation}
%
Using the inverse Fourier relations \eqref{kSpaceOpsDefn}, we obtain the following position space form of the correlation matrix: 
%
\begin{equation}
\label{CMatrixXSpace}
\langle b^{\dag}(m)b(n)\rangle=\sum_{k}\vert\beta_{k}\vert^{2}e^{2iak(m-n)};~~
\langle d^{\dag}(m)d(n)\rangle=\sum_{k}\vert\alpha_{k}\vert^{2}e^{2iak(m-n)};~~
\langle b^{\dag}(m)d(n)\rangle=\sum_{k}-\alpha_{k}\beta_{k}^{*}e^{2iak(m-n)}
\end{equation}
In the limit of large $N$, these sums transform into integrals, and we may then rewrite the correlation matrix elements \eqref{CMatrixXSpace} as:
\begin{multline}
     \label{CMatrixXSpaceLim}
\langle b^{\dag}(m)b(n)\rangle=\frac{1}{2\pi}\int_{-\pi }^{\pi } \frac{\sin ^2\left(\frac{x}{2}\right) e^{i(m-n)x}}{D^2 + 
\sin^2\left(\frac{x}{2}\right)} \, dx;~~
\langle d^{\dag}(m)d(n)\rangle=\frac{1}{2\pi}\int_{-\pi }^{\pi } \frac{D^2 \, e^{i(m-n)x}}{D^2+\sin ^2\left(\frac{x}{2}\right) } \, dx;
\\
\langle b^{\dag}(m)d(n)\rangle=\frac{1}{2\pi}\int_{-\pi }^{\pi } \frac{D \left(-i + \cot \left(\frac{x}{2}\right)\right) \sin ^2\left(\frac{x}{2}\right) e^{i(m-n)x} }{ D^2  + \sin ^2\left(\frac{x}{2}\right)} \, dx \, . \qquad \qquad
\end{multline}
We can express the $2N\times2N$ correlation matrix as a block matrix comprising four $N\times N$ blocks:
\begin{equation}
    \label{CmatrixBlocks}
    \begin{pmatrix}
    C_{11}&C_{12}\\
    C_{21}&C_{22}\\
    \end{pmatrix}
\end{equation}
where $C_{11}$ and $C_{22}$ model the $bb$ and $dd$ correlations, respectively, while $C_{12}$ and $C_{21}$ model the $bd$ correlations.  However, it is preferable to view the full correlation matrix $C$ as a block Toeplitz matrix by representing it in a form comprised of $2\times2$ matrices modeling $b$ and $d$ correlations at the {same} site in the continuum (or adjacent linked sites in the discrete model)~\cite{Widom_1974,Basor_1994}. More illustratively, we may write the correlation matrix as 
\begin{equation}
    \label{CMatrixBlockToeplitzForm}
    \begin{pmatrix}
    A_{11}&&A_{12}&&...&&A_{1N}\\
    A_{21}&&A_{22}&&...&&A_{2N}\\
    .&.&.&.\\
    .&.&.&.\\
    .&.&.&.\\
    A_{N1}&&A_{N2}&&...&&A_{NN}\\
    \end{pmatrix}
\end{equation}
where $A_{ij}$ is the $2\times2$ matrix  
$$\begin{pmatrix}
\langle b^{\dag}(i)b(j)\rangle&&\langle b^{\dag}(i)d(j) \rangle \\
\langle d^{\dag}(i)b(j)\rangle&&\langle d^{\dag}(i)d(j) \rangle
\end{pmatrix} \, .
$$
Since $A_{ij}$ depends only on $i-j$, it is clear that the correlation matrix is Toeplitz, not in its matrix elements, but in the $A_{ij}$ blocks. Mathematically, this follows from the matrix elements; physically, correlation is a sole distance spin-dependent. Note that the integrals appearing in the correlation matrix elements are simply Fourier coefficients (over the domain $S^{1}$) of the Hermitian matrix function:
\begin{equation}
  \begin{pmatrix}
  \frac{\sin ^2\left(\frac{x}{2}\right)}{D^2+\sin
   ^2\left(\frac{x}{2}\right)}
   && 
   \frac{D \left(i - \cot \left(\frac{x}{2}\right)\right)}{D^2
   \csc ^2\left(\frac{x}{2}\right)+1}
   \\[10pt]
    \frac{ D \left(-i - \cot \left(\frac{x}{2}\right)\right)}{D^2 \csc ^2\left(\frac{x}{2}\right)+1 }
   &&
  \frac{D^2}{D^2+\sin ^2\left(\frac{x}{2}\right)}
   \end{pmatrix}
\end{equation}
This observation is a common feature possessed by Toeplitz matrices appearing in similar contexts~\cite{Basor_1994,Widom_1974} and will drive most of the discussions to come.
Indeed, motivated at least partly by entanglement entropy computations, Toeplitz matrices of such kind have been studied extensively for decades leading to a wealth of information and elegant results on the spectra of such matrices in the large-$N$ limit~\cite{Basor_1994,Widom_1974}.

\section{Analytical evaluation of the entanglement entropy}
\label{sec:FermionAnalytics}
In this section, we explicitly obtain the correlation matrix by starting with single lattice point computations. Then, using the results we obtain for this special case, we draw parallels with the analogous Bosonic framework and show the regularity of the entropy in a framework imbibing zero mode contributions. As we shall see in the next section, these computations surprisingly turn out to be useful even for discussions in the large subsystem limit.

\subsection{Single lattice point computations}
\label{SmallTrace}

The simplest case we consider is where we trace over all but two of the lattice points on the staggered lattice. This corresponds to two spin degrees of freedom at one single space point in real space. We can compute the correlation matrix of such a system in terms of the parameter $K=ma$. However, since we are interested in looking at the relation between entropy and zero modes, or remnants of such relations, we express the correlation matrix in terms of the energy eigenvalues. Recall that our energy eigenvalues (which we henceforth denote by $\omega_{k}$) follow a dispersion relation of the form:
\begin{equation}
    \label{DispersionRelation}
    \omega_{k}^{2}=m^{2}+\frac{\sin^{2}(ka)}{a^{2}} \, .
\end{equation}
In the $a\rightarrow 0$ limit, this reduces to the relativistic energy-momentum dispersion relation. The correlation matrix can be shown to take the form:
\begin{equation}
\label{CrMatrix}
   \frac{1}{N} \begin{pmatrix}
    \sum_{k}\frac{\omega_{k}+m}{2\omega_{k}}&\sum_{k}e^{ika}\sqrt{\frac{\omega_{k}^2-m^2}{4\omega_{k}^2}}\\
   \sum_{k}e^{-ika}\sqrt{\frac{\omega_{k}^2-m^2}{4\omega_{k}^2}} & \sum_{k}\frac{\omega_{k}+m}{2\omega_{k}}\\
    \end{pmatrix}
\end{equation}
We contrast this with the covariance matrix obtained in Ref.~\cite{Mallayya_2014} by tracing over all, but one of the DOFs of a Hamiltonian modelling discretized free scalar field:
\begin{equation}
   \frac{1}{N} \begin{pmatrix}
    \sum_{k}\frac{1}{a\omega_{k}}&0\\
   0& \sum_{k}a\omega_{k} \, . \\
    \end{pmatrix}
\end{equation}
There are a few key differences between the two matrices above. First, the latter is diagonal, while the former is not. This is because the cross terms in the matrix model correspond to the interactions between different Weyl components at a single point. Thus, the off-diagonal terms are indicators of spin correlations. Note that coupling between different
spin components exists even in the massless limit for 
this basis choice. The chiral decoupling, analogous to the one observed in Dirac theory in $(3+1)-$D, does not occur, and the correlation matrix is still non-diagonal~\textcolor{blue}{\cite{Susskind_1977}}. Second, the two diagonal elements in the scalar field covariance matrix are quite different, a feature not observed in its spinor analog. Third, and most importantly, the Fermionic correlators in the matrix elements \emph{do not} diverge in the $m\rightarrow 0$ limit thanks to the presence of the mass term in the numerator. In the scalar field case, the mass factor was replaced by $1/a$ in the numerator leading to a divergence in the massless limit~\cite{Mallayya_2014,2020-Chandran.Shanki-PRD}. In this case, however, it is the {IR cutoff} and \emph{not} the {UV cutoff} that multiplies the energy eigenvalues, and this prevents a divergence.

Using the matrix of correlators and Eq.~\eqref{CovMatToEntHam}, it is straightforward to calculate the entanglement entropy for this system. 
Specifically, the eigenvalues of the covariance matrix are given by
\begin{equation}
\lambda_{\pm} = \frac{1}{2} \pm \pi \, \Delta ;
\quad \Delta = \sqrt{K^2 \left[F\left(-\frac{1}{K^2}\right)  - E\left(-\frac{1}{K^2}\right)\right]^2 + F\left(-\frac{1}{K^2}\right)^2} \, ,
\end{equation}
where $F$ and $E$ denote complete elliptic functions of the first and second kind, respectively. (We desist from using the more standard $K$ for the elliptic integral of the first kind to avoid confusion with the entanglement parameter $K$). The entropy, while not pretty, can be computed by plugging these eigenvalues into the expression \eqref{CorToEE}. 

We defer a discussion of the physics of entropy to section  
 \ref{Numerics}, by which time we will also have said enough to understand better the interrelations between the above expressions and the large $N$ limit. In the rest of this section, we continue with the analytical approach and obtain entanglement entropy in the limit of large system and subsystem sizes.

\subsection{Entanglement entropy for the full model}
\label{TraceLarge}

The main hindrance to obtaining analytic results for entanglement entropy in Fermionic systems lies in need for a well-defined \emph{position basis}. Such bases, which naturally emerge in Bosonic systems, allow us to transform the linear algebraic calculations into more nuanced, but {computationally} much easier problems involving differential operators. This is the same reason for the enhanced solvability of systems in quantum {mechanics} over quantum {field} theory. In the field theory, we are nearly always confined to abstract ket spaces instead of the usually encountered position space Schrodinger equation. Since the transit to differential equations is \emph{not possible} for Fermionic systems, we are forced to deal with rather heavy linear algebraic computations. In particular, it is evident from Eq.~\eqref{CorToEE} that the eigenvalues of a Toeplitz matrix whose size scales linearly with that of the system are required. Thus, finite $N$ computations inevitably require numerics to evaluate the entropy explicitly. 

However, in the $N\rightarrow\infty$ limit, the problem of computing individual eigenvalues is circumvented by rewriting the expression for the entropy as a contour integral involving the determinant of the Toeplitz matrix rather than its individual eigenvalues. 


It then becomes a question of finding the {determinant} of said Toeplitz matrix in the $N\rightarrow\infty$ limit. Analytic expressions for {this} determinant have been obtained for multiple standard spin chain models, such as the XX model~\cite{Jin_2004} and the XY model~\cite{Its_2005}, via the reformulation of the determinant evaluation into a tractable Riemann Hilbert problem. Moreover, a generic extension to quadratic Hamiltonians was carried out in Ref.~\cite{Its_2008} and can be extended for the Dirac Hamiltonian.

To do that, we rewrite the Hamiltonian \eqref{FermHam1+1DLatSussFull} in a form that will allow us to map our Hamiltonian to a well-known analog. We do this in XY steps. First, note that the transformation $b(n)\rightarrow b^{\dag}(n), d(n) \rightarrow d(n)$ does not affect the canonical anticommutation relations.   Substituting these in the Hamiltonian \eqref{FermHam1+1DLatSussFull} leads to:
\begin{equation}
    \label{HamInt1}
    H=\sum_{n=1}^{N} \left[\left(\frac{i}{2a}(b(n)d(n)-d^{\dag}(n)b^{\dag}(n)\right)-m(b^{\dag}(n)b(n)+d^{\dag}(n)d(n))\right]
\end{equation}
upto an inconsequential zero point energy. Second, we canonically transform $b(n)$ to $i b(n)$. Finally, noting that the entanglement entropy being dimensionless is independent of the scale of the problem, we rescale the Hamiltonian by a factor of $2/m$, we have:
\begin{equation}
    \label{HamInt2}
H_{r}=\sum_{n=1}^{N}\left[\left(\frac{1}{K}(b^{\dag}(n)d^{\dag}(n)-b(n)d(n)\right)-2(b^{\dag}(n)b(n)+d^{\dag}(n)d(n))\right]
\end{equation}
which, on reverting to the staggered lattice variable $\phi(n)$ yields
\begin{equation}
    \label{HamFin}
    H_{r}=\sum_{n=1}^{2N} \left[\left(\frac{1}{K}(\phi^{\dag}(n)\phi^{\dag}(n+1)-\phi(n)\phi(n+1)\right)-2(\phi^{\dag}(n)\phi(n)\right]
\end{equation}
where $K\equiv m a$ is a dimensionless parameter balancing the contributions of the IR and UV cutoffs of the system.  

Before moving forward with the rest of the analysis, we want to provide an understanding of the origins of the above transformations. First, we have precedent for renaming a field operator (i.e., $\Psi\rightarrow\Psi^{\dag}$). This is a crucial step of the canonical quantization of the Dirac field in $(3+1)-$D, which allows one to obtain a manifestly positive Hamiltonian at the cost of a divergent zero point energy~\cite{Peskin_1995}. Thus, it is no surprise that a lower dimensional analog appears in our discrete model. Second, the multiplication of the $b$s with an $i$ essentially corresponds to choosing a different convention for the anticommuting matrices appearing in the continuum theory. Such a switch is expected as the initial choice was mostly motivated by reasons of formality, and apriori had no reason to be relevant for entropy computations. However, the question of locality must be considered, as while \textit{global} basis changes are allowed, spacetime-dependent choices require us to gauge the Dirac field. Indeed, by construction, variables on the odd and even lattice sites transform differently. Although the transformation is local in the {Susskind} lattice (since the $b$s and the $d$s transform differently), it is {global} in the continuum wherein neighboring $b$s, and $d$s coalesce into components of a 2-spinor localized to a \textit{single} space point. The transformation we have imposed is {effectively} a global transformation, not a gauge transformation. Other global transformations are possible; we picked a concrete example from the mix. 

The resulting Hamiltonian \eqref{HamFin} is, at least functionally, well-known in condensed matter literature \cite{Latorre_2009}. Indeed, a hopping model with an identical structure and parametric dependence can be obtained starting from the Hamiltonian for the ferromagnetic XY chain~{\cite{Latorre:2003kg}}
\begin{equation}
    \label{XYHam}
    H_{XY}\equiv-\frac{\alpha}{2}\sum_{i=1}^{N}\left[(1+\gamma)\sigma^{x}_{i}\sigma^{x}_{i+1}+(1-\gamma)\sigma^{y}_{i}\sigma^{y}_{i+1}\right]-\sum_{i=1}^{N}\sigma^{z}_{i}
\end{equation}
Via the Jordan-Wigner transform, the above Hamiltonian can be mapped to a free-hopping model~{\cite{Latorre:2003kg}}:
\begin{equation}
\label{XYHamJW}
H_{XY}^{JW}=\sum_{n=1}^{N}\left[(\frac{\alpha}{2}(\phi^{\dag}(n)\phi(n+1)+\phi^{\dag}(n+1)\phi(n)+\gamma(\phi^{\dag}(n)\phi^{\dag}(n+1)-\phi(n)\phi(n+1))-2(\phi^{\dag}(n)\phi(n)\right]
\end{equation}
The constant $\alpha$ quantifies the strength of an applied magnetic field, while $\gamma$ parameterizes the anisotropy. The similarities between the hopping model above and that of the discretized Dirac Hamiltonian \eqref{HamFin} are now apparent. In particular, the latter is obtained in the bivariate limit $\alpha\rightarrow 0, \gamma\rightarrow \infty, \gamma\alpha\rightarrow\frac{2}{K}$. Note that the requirement of ferromagnetism constrains the parameter $\gamma$. Hence, for the ferromagnetic case, the parameters $(\alpha, \gamma)$ in $\eqref{XYHam}$ must be positive (and negative for the antiferromagnetic case). The conditions we must impose to obtain the discretized Dirac Hamiltonian as a limiting case certainly break these requirements. While it is {mathematically} consistent (at least so long as a well-defined ground state exists), the {physical} relationship between the two setups --- a spinorial field on the one hand and a spin chain on the other --- is rather non-trivial. The former corresponds to a ferromagnetic chain along one direction and antiferromagnetic along another. However, there is no reason {not} to take the limit; in particular, if a given mathematical statement holds for a generic range of $\gamma$ in the XY analysis, we may carry it over to our problem. Dealing with constrained statements that apply for $\gamma\in\{0,1\}$ is more nuanced; however, as we shall see, such situations are manageable when they arise.
Nevertheless, the physics of the $XY$ chain appears to be fundamentally distinct from that of the spinorial field; consequently, it is not appropriate to pull final results directly from the $XY$ model.

After this digression, let us obtain the entanglement entropy in the continuum limit. Following Ref.~\cite{Its_2008}, we see that the entanglement for general translationally invariant spin chains is captured by a single polynomial $p(z)$ of degree $2N$ with coefficients depending algebraically on the couplings of the Hamiltonian. The dependence of the entanglement entropy on a set of parameters growing polynomially with the lattice size $N$ is seemingly at odds with the exponential growth of the Hilbert space with $N$. However, the entanglement entropy depends not just on the structure of the underlying Hilbert space but also the state in question and its partition into subsystems. In particular, the growth of correlations with $N$ is a better measure for quantifying the corresponding growth of the entropy. As quantified by the correlation matrix, the number of independent correlation functions grows polynomially with $N$.

The algebraic equality of \eqref{HamFin} with a limiting version of \eqref{XYHamJW} allows us to work with the polynomial directly derived in Ref.~\cite{Its_2005,Its_2008} for the $XY$ model. Thus, we can obtain the relevant polynomial (prior to taking the bivariate limit) as
\begin{equation}
    \label{ItsPol}
    p(z)=\frac{\alpha(1-\gamma)}{2}z^{2}-z+\frac{\alpha(1+\gamma)}{2} \, .
\end{equation}
In the limiting case, we obtain the following $K$ parametrized family of polynomials:
\begin{equation}
    \label{OurPol2}
  p(z,K)=-\frac{1}{K}z^{2}-z+\frac{1}{K}  \, ,
\end{equation}
which, in principle, carry all information on the entanglement entropy and divergences therein. A geometric reformulation of the problem, outlined in Ref.~\cite{Its_2008}, posits entanglement entropy divergences whenever pairs of roots of the polynomial $q(z, K)\equiv z^{N}p(z, K)p(1/z, K)$ approach the unit circle $S^{1}$ on the complex plane. More precisely, the reality of the coefficients of $p(z)$ and the form of $q(z)$ together imply that roots of $q(z)$ come in conjugate reciprocal pairs, i.e., $\overline{\zeta_{i}^{-1}}$ is a root of $q(z)$ if $\zeta$ is. If a subset $\{\zeta_{i}\}, i=1,...,m$ of roots of $q(z)$ confined to the interior of $S^{1}$ approach its boundary on varying $K$, then it is clear that the subset $\{\overline{\zeta_{i}^{-1}}\}$ approach $S^{1}$ from its exterior. From the geometric viewpoint, in this limit the entanglement entropy diverges as:
\begin{equation}
\label{GeomDiv}
S\sim - \frac{1}{6} \sum_{i=1}^{m} \log(\vert\zeta_{i} -\overline{\zeta_{i}^{-1}} \vert)
\end{equation}
In our specific instance, the roots of the polynomial $p(z)$ are given by 
\begin{equation}
    \label{RootsP}
    \zeta_{\pm}=\frac{-K\pm\sqrt{K^{2}+4}}{2}
\end{equation}
tracing through the steps outlined in the above paragraph, we find that the entanglement entropy diverges in the $K\rightarrow 0$ limit (and \textit{only} in this limit)  as
\begin{equation}
    \label{EEDiv}
    S\rightarrow -\frac{1}{3}\log(\vert K\vert)
\end{equation}
We thus again observe that the emergence of zero modes ties in with the divergence of the entanglement entropy. Nevertheless, there are significant differences between these ties and those we observed for Bosonic chains. In the case of Fermions, divergences emerged only when \textit{both} the system \textit{and} the subsystem size ceased to be finite. However, divergences occur, even for a finite number of Bosonic chains.

Of course, it is no surprise that finite Fermionic systems yielded no divergences - their Hilbert spaces are finite-dimensional. That said, it would be desirable to have a single unified framework for predicting the occurrence of divergences: one which could encompass both Bosonic and Fermionic systems and not rely on any `external assistance' (e.g., the finiteness of Fermionic Hilbert spaces). We already have such a framework for Bosonic chains; in fact, we have explored two distinct sets of signatures, one involving the zero modes of the system Hamiltonian and the other involving the \textit{entanglement} Hamiltonian. The former was slightly on the heuristic side but was nevertheless instructive, while the latter was quite precise. As it turns out, \textit{both} of these methodologies \textit{can} be extended to the Fermionic sector. 

The key theme underlying the first of these extensions is a shift in perspective from the zero modes themselves to the \textit{tower of states generated by these modes}. Explicitly, the Pauli exclusion principle implies that the ladder of states generated by a single Fermionic zero mode has just a \textit{single} rung. In contrast, a single Bosonic zero mode is associated with an infinite tower of states generated by the corresponding raising operator. As we had seen in Sec \eqref{sec:CHO}, the presence of a closely separated tower of near zero states was a necessary and sufficient requirement for entropy divergences. With this interpretation, the finiteness of entropy for finite Fermionic systems with zero modes can be attributed to the inability of these modes to generate an \textit{infinite tower} of closely separated energy states.
While the discretized Dirac Hamiltonian contains just a single mode regardless of its lattice size, there is an evident increase in the number of \textit{near-zero modes} as we approach the limit of infinite lattice size. This follows from the discrete dispersion relation \eqref{DispersionRelation}. Even though {individual} near zero modes can only generate single-rung towers of near-zero eigenstates when the vacuum acts on their corresponding ladder operators, the large $N$ limit contains an ever-increasing number of near-zero eigenstates, leading to entropic divergence. This is indeed what we observed earlier!
The second class of signature of diverging entropies concerned the \textit{entanglement} Hamiltonian and its zero modes, in particular. 

The formal extension of this statement to Fermionic systems is trivial --- following such an extension, we would claim that entropic divergences must necessarily be accompanied by the emergence of zero modes of the entanglement Hamiltonian. Moreover, we may show that such zero modes translate to an eigenvalue of $1/2$ for the correlation matrix~ \cite{Peschel_2003}. Indeed, we find that the finite-dimensional correlation matrix \eqref{CrMatrix} does not have such eigenvalues, so divergences of the entropy cannot occur in the finite lattice regime. On the other hand, numerical enumerations of the eigenvalues of the correlation matrix in $N\rightarrow\infty$ limit indicate that $1/2$ serves as an accumulation point in the massless case, so our methodology does seem adequate.
Having analytically identified a pair of coherent, self-contained frameworks, which can correctly signal entropic divergences for both Bosonic and Fermionic systems, we now turn our focus to numerics in the following section.

\section{Comparison of analytical expressions with numerics}
\label{Numerics}

This section provides a repository of the numerical results that corroborate the analytic computations of the past few sections. 

\begin{figure}[!htb]
\centering        \includegraphics[width=0.75\textwidth]{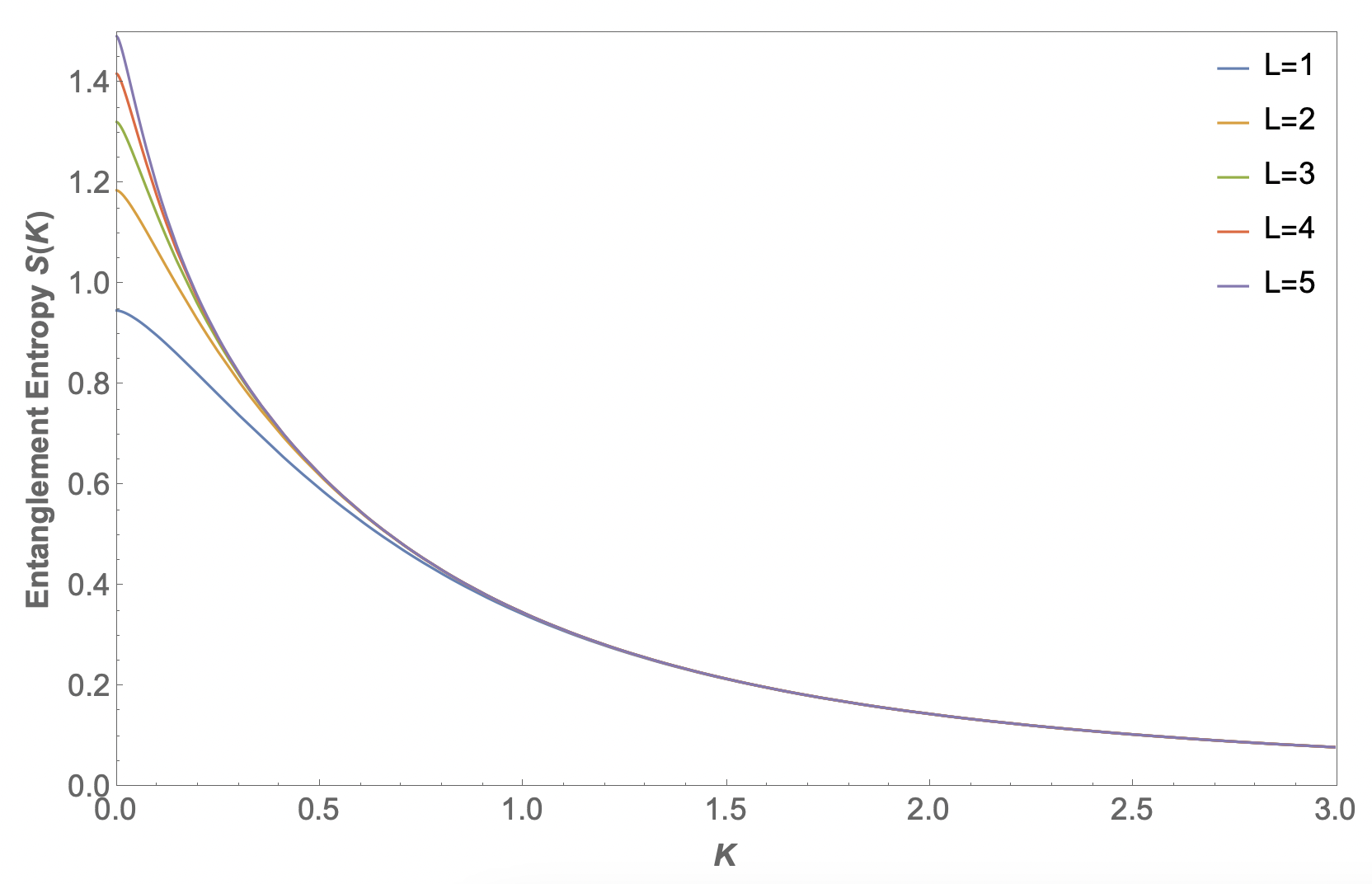}
\caption{Entanglement Entropy $(S)$ as a function of $K$. The full chain is taken to be very large.}
     \label{EEvsK}
     \hfill
\end{figure}

\ref{EEvsK} presents the results of numerical evaluation of the entanglement entropy of the ground state of the discretized Dirac Hamiltonian \eqref{HamInt1} in the small subsystem limit. Again, the entanglement entropy is invariant under an overall scale transformation and depends solely on the parameter $K=ma$. We infer the following from \ref{EEvsK}:
First, as expected, for a fixed $K$ value, the entropy increases with an increase in subsystem size and saturates to a finite value in the $K\rightarrow 0$ limit with the Fermionic nature of the system ensuring the finiteness of the entropy. Second, exact computations in this limit are indeed possible, with the computations carried out in \ref{SmallTrace} for the $L=1$ case perfectly matching the numerical computations. 

Lastly, it is worth observing the rapid destruction of $L$ dependence. For $K$ values as small as $3$, the entanglement entropy is insensitive to the subsystem size $L$. This, of course, suggests a saturation 
in entropy with subsystem size. Thus, the small subsystem provides us with an arena where simple analytic computations bear fruit and {also} surprisingly predict a generic trend for the large subsystem limit as well --- that the entropy asymptotically saturates, at least for "large enough" $K$. Of course, since there should be no \textit{qualitative} difference for different \textit{non-zero} values of $K$, the small subsystem numerics tell us that the entropy should saturate for {any} non-zero $K$. Indeed, this is what one expects for large $L$ computations, given that we obtained divergence strictly in the massless limit, with no middle ground between divergence and saturation. It is quite interesting that one can obtain a very good understanding of the large subsystem limit even from working on a completely different regime.

\begin{figure}[!htb]
\begin{center}
\subfloat[\label{sym1a}][]{%
\includegraphics[width=0.5\textwidth]{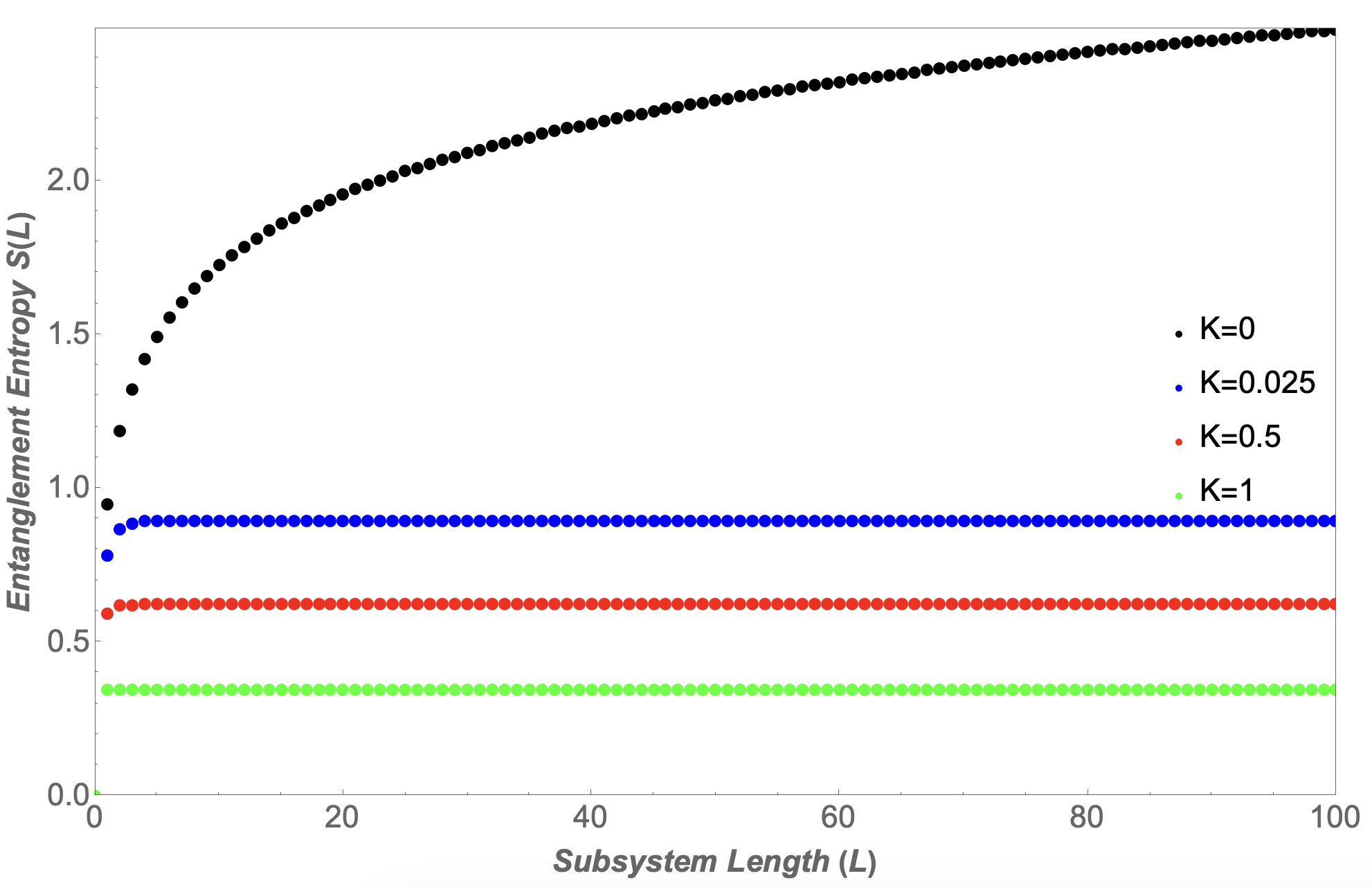}}
\subfloat[\label{sym1b}][]{			\includegraphics[width=0.5\textwidth]{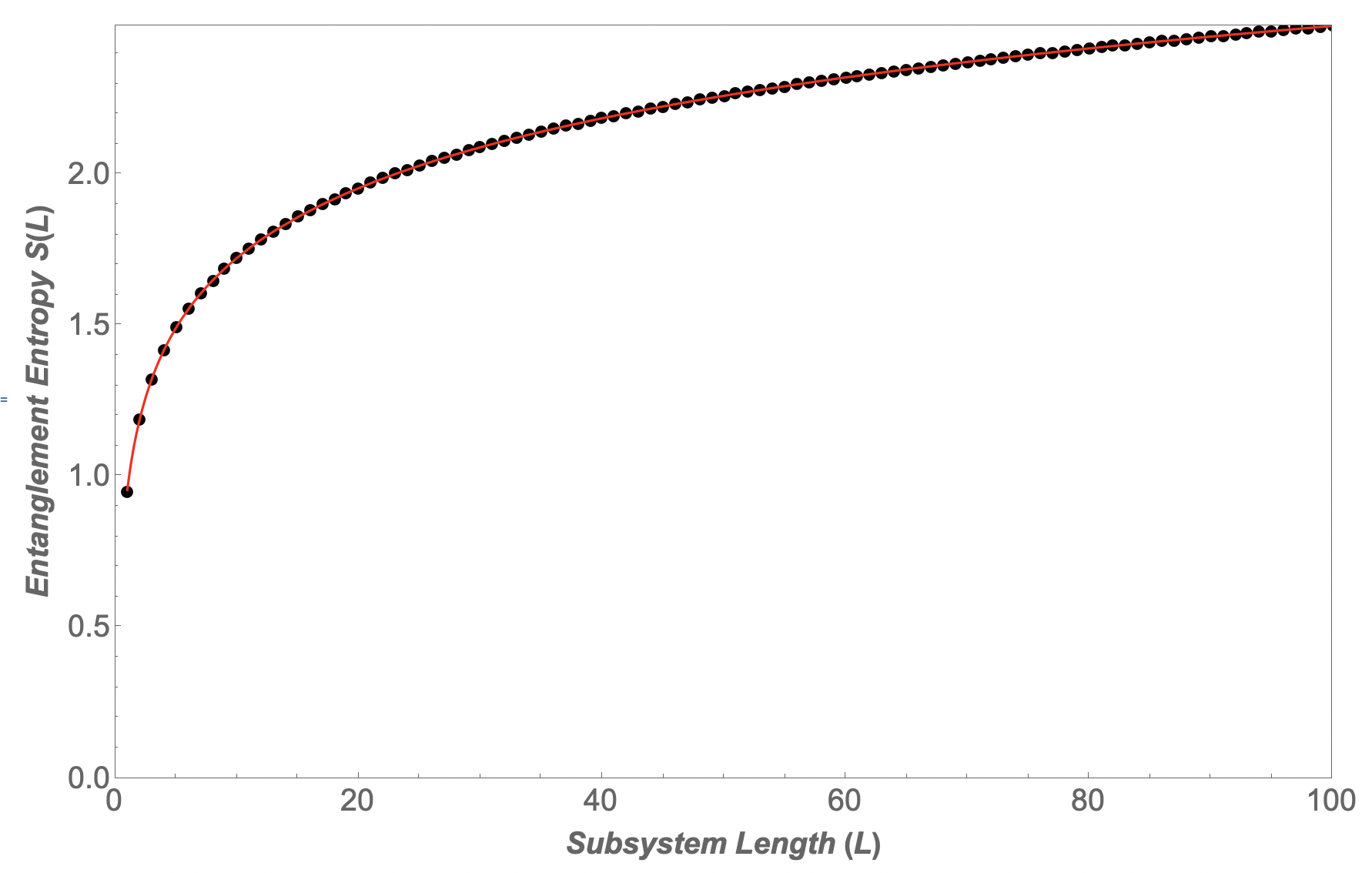}}
\caption{(a) Entanglement Entropy ($S$) as a function of Subsystem size for different values of $K$. (b) Entanglement Entropy ($S$) as a function of Subsystem size in the zero $K$ limit. {The full chain is taken to be very large.}}
        \label{EEvsL}
     \hfill
 \end{center}    
\end{figure}

Having discussed the physics of large chains from a completely different direction, we now move on with \textit{genuine} large $L$ computations. As evidenced from the left plot in \ref{EEvsL}, simple numerics are enough to demonstrate the saturation of the entropy even for $K$ values rather close to 0. The $K=0$ behavior is markedly different from its counterparts, with no signs of saturation even for subsystem sizes of order $\mathcal{O}(100)$ and an apparent logarithmic trend. Indeed, logarithmic fits provide good estimates of the entropy (right plot of \ref{EEvsL}), with the entropy going as 
\begin{equation}
    \label{EEEstimate}
    S(K)\sim A\log(K)+B \, ,
\end{equation}
where the parameter values are numerically found to be $A\sim0.3338$ and $B\sim0.9553$. Thus, the logarithmic prefactor of $1/3$, obtained via the calculations in section \ref{TraceLarge} are reproduced accurately by the numerics, with the prefactor stemming from the central charge of the associated conformal field theory~\cite{Castro_Alvaredo_2009}.

\begin{figure}[!htb]
\begin{center}
\subfloat[\label{sym1aI}][]{%
\includegraphics[width=0.5\textwidth]{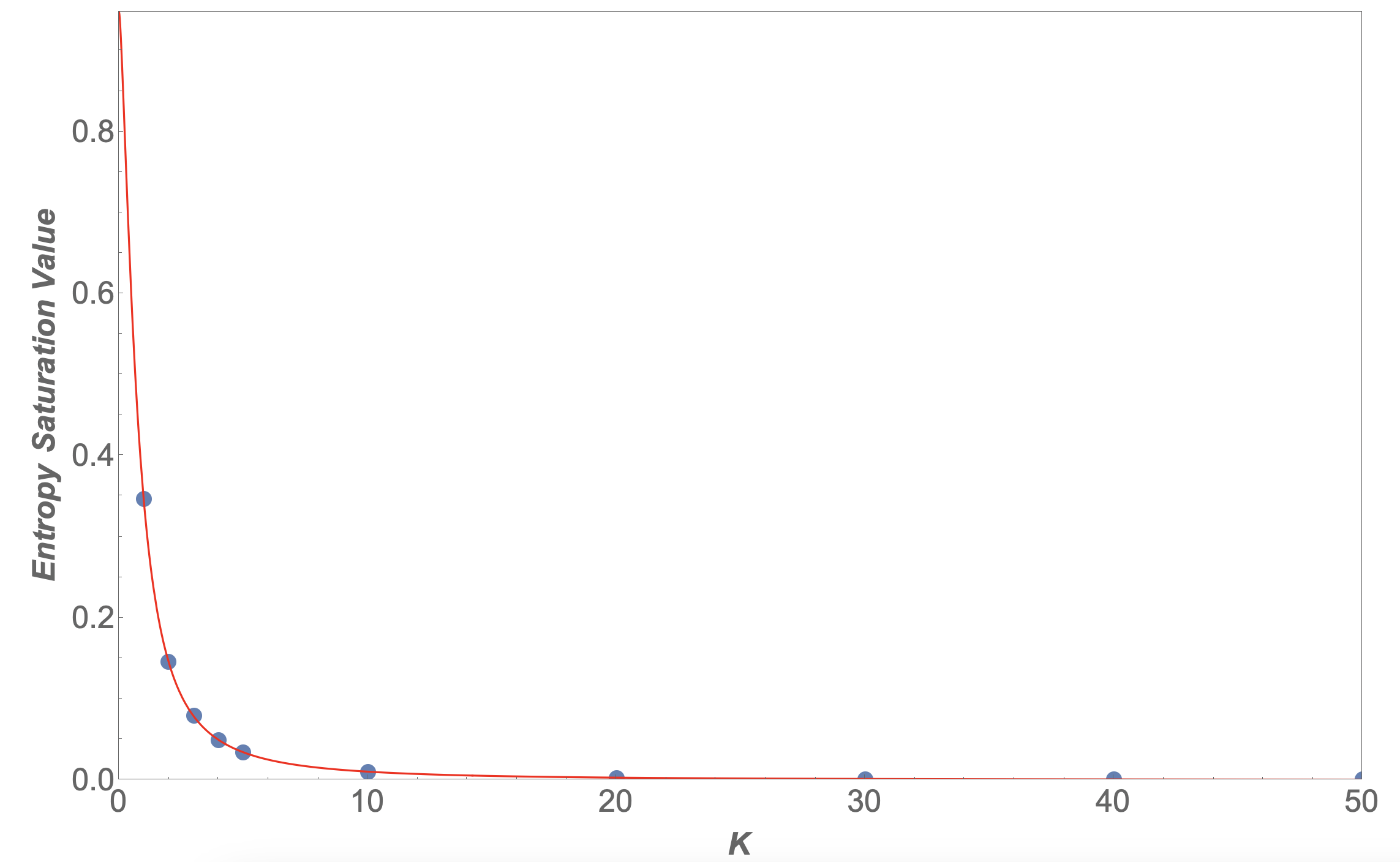}}
\subfloat[\label{sym1bI}][]{			\includegraphics[width=0.5\textwidth]{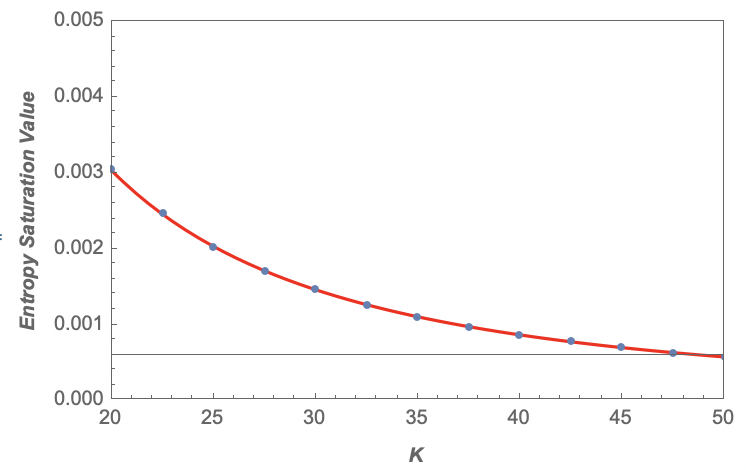}}
\caption{(a) Entropy Saturation Value as a Function of $K$ for the range $0 < K < 50$. (b) Entropy Saturation Value as a Function of $K$ for the range $20 < K < 50$. {The full chain is taken to be very large.}}
\label{SatFit1}
\hfill
\end{center}
\end{figure}

This leads us to the following question: What of the saturation values? Here, our numerics \textit{really} come in handy as we do not have a clear-cut way to estimate the entropy (and thus its limiting value) for massive systems. The aforementioned rapid loss of sensitivity to subsystem size means that we may use the single lattice point computations to obtain good estimates for the saturation values of the entanglement entropy for arbitrary $K$ values. Indeed, fitting the computed saturation values for an appropriately chosen set of $K$ values using the $L=1$ curve in \ref{EEvsL} yields excellent results. The left plot of \ref{SatFit1} shows these fits at a gross level. That said, precise results can be obtained at a more fine-grained level, as in the right plot of \ref{SatFit1}, where we focus on larger $K$ values. Thus, we see that the small subsystem computations provide accurate predictions of the large $L$ saturation points, even when the numerics are not too sophisticated --- $K$ values large and the saturation values tiny. Once again, we take note of the quickly emerging quantitative similarities between the two limits ---large $L$ and small $L$.
%


\section{Conclusions and Discussions}
 
This work probed the origins of divergent entanglement entropy for Bosonic and Fermionic chains. Starting with a system comprising two Bosonic degrees of freedom, we noted the connections between divergent entropy and the phenomenon of emergent zero-modes. We supplanted this with a set of heuristically motivated criteria for signaling divergences. Generalizing to $N$-Bosonic chains, we then altered our perspective of the entanglement divergence to zeros of the \textit{entanglement} Hamiltonian. In so doing, we derived an additional, more quantitively precise diagnostic for diverging entropy. 

Shifting to Fermionic systems and motivated by field theory, we considered the example of a discretized chain of staggered fermions in one dimension. First, we analytically deduced the existence of a diverging entropy in the infinite subsystem limit. We then combined our analytical knowledge with our prior results for Bosonic chains to obtain a completely inclusive set of diagnostics for divergent entropies, encompassing both the Fermionic and Bosonic regimes. Specifically, we found that the emergence of zero modes, or more generically, near-zero modes, of the \textit{entanglement} Hamiltonian was a necessary and sufficient criterion for obtaining divergent entanglement entropy. In contrast, zero-modes of the system Hamiltonian are a necessary, but not sufficient condition for entropy divergences. However, stronger statements can be made by observing the tower of zero or near-zero eigenstates that the corresponding ladder operators generate. Finally, we confirmed the approximate analytical results with numerics.

An attentive reader might ask: Whether the factorization algebra theorem fails in an infinite dimensional Hilbert space system? How do zero-modes fit in the factorization algebra theorem?  
The current analysis can not make any affirmative statement about the validity of the factorization algebra theorem for infinite dimensional Hilbert space systems. However, our analysis suggests that the factorization algebra theorem can not apply to systems with zero modes. 

Zero-mode states increase the Hilbert space of the system but do not contribute to the system's energy. In other words, a countably infinite number of zero-modes do not contribute to the Hamiltonian. However, if we remove a few of the zero mode states, then effectively, the system is not in a pure state. Moreover, as mentioned in the introduction, a maximally mixed state is separable for any factorization. Therefore, different factorizations can lead to maximally divergent entropy, and some can lead to finite entropy. This is analogous to different paths in the phase space (cf. \ref{ParameterSpacePaths}) that lead to wildly different values for entanglement entropy.

The above results bring attention to the following interesting questions: First, while the criteria we outlined were enough to explain the divergence structure of all the systems we considered, they require more rigorous formulations. For instance, while we established that a certain `proximity' was required between near-zero eigenstates for generating divergences in the entropy, quantitative notions of this proximity are not obvious. Additionally, it is unclear exactly how this tower of states alters the ground state leading to divergent subsystem entropy. Similarly, a physical understanding of the zero modes of the entanglement Hamiltonian and their relation to the divergences is still required.

Second, given that our studies were aimed at studying entropy divergences in Quantum Field Theory (QFT), it is crucial to understand how these findings apply to continuum theories. Lastly, other ideas involve studying the validity of these conditions for higher dimensional systems. For example, it is well known that an increase in dimension softens the effects of infrared divergences, thereby downplaying the effects of zero modes. It would thus be interesting to see how the themes we have outlined in this article pan out in higher dimensions.

\section{Acknowledgements}
The authors thank S. Mahesh Chandran for his comments on the earlier version of the manuscript. Thanks to A. Kushwaha for helping with the Tikz plot. VN acknowledges the support of the Indian Institute of Technology, Bombay for this work. The work is supported by the SERB-MATRICS grant. 

\appendix
\section{Spectrum of the discretized Hamiltonian~\eqref{FermHam1+1DLatSussFull}}
\label{app:spectrum}

In this appendix, we compute the spectrum and the eigenstates of the discretized Hamiltonian~\eqref{FermHam1+1DLatSussFull}, which is a natural step in understanding any quantum system. Here, we focus on the spectrum from the zero modes point of view. 

The factor of $(-1)^{i}$ in \eqref{FermHam1+1DLatSussFull} is disconcerting. So we begin our study of the massive Hamiltonian by rewriting \eqref{FermHam1+1DLatSussFull} in a more explicit form, distinguishing between the individual components of the Weyl spinors. We have 
\begin{equation}
\label{FermHam1+1DLatSussFullExp}
     \sum_{i=1}^{N}[\frac{i}{2a}(b^{\dag}(n)d(n)-d^{\dag}(n)b(n))+m(b^{\dag}(n)b(n)-d^{\dag}(n)d(n))]
\end{equation}
where $b$s and $d$s respectively model, in the discrete regime, the two annihilation operators correspond to the continuum Weyl field. The CCR \eqref{CCRSuss} is then rewritten as
\begin{equation}
    \label{CCRExp}
    \{b(n),b^{\dag}(m)\}=\{d(n),d^{\dag}(m)\}=\delta_{nm}
\end{equation}
with all other anti-commutators yielding zero. Since the Hamiltonian \eqref{FermHam1+1DLatSussFull} is invariant under lattice translations of magnitude $2a$, we can solve for energy eigenstates using Fourier methods. Specifically, we define the momentum space operators $\tilde{b}(k)$ and $\tilde{d}(k)$ as
\begin{equation}
\label{kSpaceOpsDefn}
 \Tilde{b}(k)=\frac{1}{\sqrt{N}}\sum_{l}e^{-i2kla}b(l); \quad 
     \Tilde{d}(k)=\frac{1}{\sqrt{N}}\sum_{l}e^{-i2kla}d(l)
\end{equation}
where $k$ takes values from the set $\pi n/Na$, $n\in\{-(N-1)/2,2,3...(N-1)/2\}$ (we henceforth assume $N$ to be even). These relations are easily inverted to give
\begin{equation}
\label{kSpacetoxSpace}
    b(l)=\frac{1}{\sqrt{N}}\sum_{l}e^{i2kla}\tilde{b}(k); \quad 
    d(l)=\frac{1}{\sqrt{N}}\sum_{l}e^{i2kla}\tilde{d}(k)
     \\
\end{equation}
The momentum space operators are, of course, canonically conjugate as well
\begin{equation}
    \label{CCRExpkSpace}
    \{\tilde{b}(k),\tilde{b}^{\dag}(k')\}=\{\tilde{d}(k),\tilde{d}^{\dag}(k')\}=\delta_{kk'}
\end{equation}
with all other anti-commutators yielding zero. Rewriting the Hamiltonian in terms of the momentum space operators, we observe a decoupling of momentum modes so that the Hamiltonian can be expressed as
\begin{equation}
    H=\sum_{k}\Tilde{H}(k)
\end{equation}
where the momentum space Hamiltonian, $\Tilde{H}(k)$ is defined as 
\begin{equation}
\label{HamkSpace}
    \Tilde{H}(k)=
    \begin{pmatrix}\tilde{b}^{\dag}(k)&\tilde{d}^{\dag}(k)\\
    \end{pmatrix}
    \begin{pmatrix}
    m&\frac{i}{2a}(1-e^{-i2ka})\\
    -\frac{i}{2a}(1-e^{2ika})&m\\
    \end{pmatrix}
    \begin{pmatrix}
    \tilde{b}(k)\\
    \tilde{d}(k)\\
    \end{pmatrix}
\end{equation}
To diagonalize the Hamiltonian, we need to do one more rotation, this time entirely in the momentum space. Let 
$$U(k)=\begin{pmatrix}\alpha_{k}&\beta_{k}\\-\beta_{k}^{*}&\alpha_{k}^{*}\end{pmatrix} \, ,$$ 
where, $\vert\alpha_{k}\vert^{2}+\vert\beta_{k}\vert^{2}=1$ and $U(k)$  is the $SU(2)$ matrix that diagonalizes $H(k)$. Defining the canonically conjugate operators $\tilde{a}(k)$ and $\tilde{c}(k)$ as
\begin{equation}
    \label{RotOpskSpace}
    \begin{pmatrix}
    \tilde{a}(k)\\\tilde{c}(k)
    \end{pmatrix}
    =U(k)
    \begin{pmatrix}
    \tilde{b}(k)\\\tilde{d}(k)
    \end{pmatrix}
\end{equation}
we then obtain the diagonal form of the Hamiltonian as
\begin{equation}
    \label{HamDiag}
    H=\sum_{k}(E_{k}(\tilde{a}^{\dag}(k)\tilde{a}(k)-\tilde{c}^{\dag}(k)\tilde{c}(k)))
\end{equation}
where, $E_{k}=(\sqrt{\sin^{2}(k\pi/a)+m^{2}})/a$ and $-E_{k}$ are the eigenvalues of $H_{k}$. Note that, in the continuum limit, these match the Dirac dispersion relation. Thus, our computations have passed their natural check.

Contrary to usual second quantized space procedures, we do {not} define the vacuum $\vert0\rangle$ as the state annihilated by the $a$s and the $c$s. If we were to do so, the $a$s would serve as lowering operators, but the $c$s would serve as {raising} operators owing to the sign of the coefficients of the $c$ terms in $\eqref{HamDiag}$. Nevertheless, our approach is barely affected by this subtlety. We use the CCR to rewrite our Hamiltonian as 
\begin{equation}
    \label{HamDiag2}
      H=\sum_{k}(E_{k}(\tilde{a}^{\dag}(k)\tilde{a}(k)+\tilde{c}(k)\tilde{c}^{\dag}(k)))-\sum_{k}E_{k}
\end{equation}
which now has the standard form of a sum of a set of positive definite number operators weighted by positive coefficients. The only price is the introduction of a negative constant which manifests physically as a negative zero point energy (ZPE). Note that similar steps need to be performed when quantizing the continuum Hamiltonian, with the zero point energy diverging and requiring an ``infinite" shift of the Hamiltonian operator. In particular, as in the unique case of spinor fields, we obtain the well-known spinorial feature of \textit{negative} rather than \textit{positive} ZPEs~\cite{Srednicki_2007}. Our negative constant also diverges in the continuum limit. 

We now define the vacuum $\vert0\rangle$ as the unique state annihilated by the $a$s and the ${c}^{\dag}$. $a^{\dag}$s and $c$s serve as raising operators (creation operators in the continuum picture). The Hilbert space is spanned by the tower of states obtained from the vacuum. The energy of a state of the form $\Pi_{i}\Pi_{j}(\tilde{a}(k_{i})^{\dag}\tilde{c}(k_{j})\vert0\rangle$ is (after shifting to account for the ZPE) $\sum_{i}E_{k_{i}}+\sum_{j}E_{k_{j}}$ and models a system of freely hopping fermions. 
\newline

%

\end{document}